\documentclass[12pt]{article}

\usepackage{amssymb}
\usepackage{amsmath}
\usepackage{amscd}
\usepackage{latexsym}
\usepackage{graphicx}

\usepackage{cite}

\newcommand{\be}{\begin{equation}}
\newcommand{\ee}{\end{equation}}
\newcommand{\Dlt}{\Delta}
\newcommand{\dlt}{\delta}
\newcommand{\prt}{\partial}
\newcommand{\br}{{\bf r}}
\newcommand{\ba}{{\bf a}}
\newcommand{\bfe}{{\bf e}}

\newcommand{\bS}{{\bf S}}

\newcommand{\bB}{{\bf B}}
\newcommand{\bt}{\beta}
\newcommand{\vp}{\varphi}
\newcommand{\ep}{\varepsilon}
\newcommand{\al}{\alpha}
\newcommand{\ra}{\rightarrow}
\newcommand{\sgm}{\sigma}

\newcommand{\gm}{\gamma}
\newcommand{\om}{\omega}
\newcommand{\Om}{\Omega}

\newcommand{\dgr}{\dagger}
\newcommand{\lbd}{\lambda}
\newcommand{\Lbd}{\Lambda}
\newcommand{\rgl}{\rangle}
\newcommand{\lgl}{\langle}
\newcommand{\cH}{{\cal H}}

\begin{document}

\begin{center}

{\Large{\bf Trapped Bose-Einstein condensates with nonlinear coherent modes} \\ [5mm]

V.I. Yukalov$^{1,2}$, E.P. Yukalova$^{3}$ and V.S. Bagnato$^2$}  \\ [3mm]

{\it
$^1$Bogolubov Laboratory of Theoretical Physics, \\
Joint Institute for Nuclear Research, Dubna 141980, Russia \\ [2mm]

$^2$Instituto de Fisica de S\~ao Carlos, Universidade de S\~ao Paulo, \\
CP 369, S\~ao Carlos 13560-970, S\~ao Paulo, Brazil \\ [2mm]

$^3$Laboratory of Information Technologies, \\
Joint Institute for Nuclear Research, Dubna 141980, Russia } \\ [3mm]

{\bf E-mails}: {\it yukalov@theor.jinr.ru}, ~~ {\it yukalova@theor.jinr.ru} \\
{\it vander@ifsc.usp.br} 

\end{center}

\vskip 1cm

\begin{abstract}

The review presents the methods of generation of nonlinear coherent excitations in 
strongly nonequilibrium Bose-condensed systems of trapped atoms and their properties. 
Non-ground-state Bose-Einstein condensates are represented by nonlinear coherent modes. 
The principal difference of nonlinear coherent modes from linear collective excitations 
is emphasized. Methods of generating nonlinear modes and the properties of the latter 
are described. Matter-wave interferometry with coherent modes is discussed, including 
such effects as interference patterns, internal Josephson current, Rabi oscillations, 
Ramsey fringes, harmonic generation, and parametric conversion. Dynamic transition 
between mode-locked and mode-unlocked regimes is shown to be analogous to a phase 
transition. Atomic squeezing and entanglement in a lattice of condensed atomic clouds 
with coherent modes are considered. Nonequilibrium states of trapped Bose-condensed 
systems, starting from weakly nonequilibrium state, vortex state, vortex turbulence, 
droplet or grain turbulence, and wave turbulence, are classified by means of effective 
Fresnel and Mach numbers. The inverse Kibble-Zurek scenario is described. A method for 
the formation of directed beams from atom lasers is reported.

\end{abstract}

\vskip 2mm

{\it Keywords}: non-ground-state Bose-Einstein condensates, coherent modes, resonant 
generation, resonant entanglement, atomic squeezing, quantum turbulence, atom laser

\newpage

{\large{\bf Contents}}

\vskip 5mm
{\bf 1. Introduction}

\vskip 2mm
{\bf 2. Condensate wave function}

\vskip 2mm
{\bf 3. Coherent modes}

\vskip 2mm
{\bf 4. Resonant generation}

\vskip 2mm
{\bf 5. Counterflow instability}

\vskip 2mm
{\bf 6. Energy levels}

\vskip 2mm
{\bf 7. Mode dynamics}

\vskip 2mm
{\bf 8. Mode locking}

\vskip 2mm
{\bf 9. Shape-conservation criterion}

\vskip 2mm
{\bf 10. Multiple mode generation}

\vskip 2mm
{\bf 11. Matter-wave interferometry}

\hspace{6mm}   11.1. Interference patterns

\hspace{6mm}   11.2. Internal Josephson current

\hspace{6mm}   11.3. Rabi oscillations

\hspace{6mm}   11.4. Higher-order resonances

\vskip 2mm
{\bf 12. Ramsey fringes}   

\vskip 2mm
{\bf 13. Interaction modulation}

\vskip 2mm
{\bf 14. Strong interactions and noise}

\vskip 2mm
{\bf 15. Critical phenomena}

\vskip 2mm
{\bf 16. Atomic squeezing}

\vskip 2mm
{\bf 17. Cloud entanglement}

\hspace{6mm}   17.1. Entangled states

\hspace{6mm}   17.2. Entanglement production

\hspace{6mm}   17.3. Resonant entanglement production

\vskip 2mm
{\bf 18. Nonresonant mode generation}

\vskip 2mm
{\bf 19. Nonequilibrium states}

\hspace{6mm}   19.1. Weak nonequilibrium

\hspace{6mm}   19.2. Vortex embryos

\hspace{6mm}   19.3. Vortex rings

\hspace{6mm}   19.4. Vortex lines

\hspace{6mm}   19.5. Vortex turbulence

\hspace{6mm}   19.6. Droplet state

\hspace{6mm}   19.7. Wave turbulence

\vskip 2mm
{\bf 20. Classification of states}

\vskip 2mm
{\bf 21. Atom laser}

\vskip 2mm
{\bf 22. Conclusion}

\section{Introduction}

Bose-Einstein condensation of trapped dilute atoms has been of high interest
in recent years \cite{Anderson_1,Bradley_2,Davis_3} since 1995, triggering voluminous 
publications, as can be inferred from review articles \cite{Parkins_4,Dalfovo_5,
Courteille_6,Andersen_7,Yukalov_8,Bongs_9,Yukalov_10,Posazhennikova_11,Yukalov_12,
Proukakis_13,Yurovsky_14,Yukalov_15,Yukalov_16}. Equilibrium states of Bose-condensed 
systems have been studied both for atoms in traps and in optical lattices 
\cite{Morsch_17,Mosely_18,Yukalov_19,Krutitsky_20,Yukalov_21}. Elementary excitations 
have been investigated, corresponding to small deviations from the ground state,
that is, to small density oscillations. These excitations are usually described by 
considering linear deviations of density from the ground state. 

In 1997, the authors \cite{Yukalov_22} suggested that it is feasible to generate in
atomic traps strongly nonlinear excitations corresponding to non-ground-state 
Bose-Einstein condensates. Since the state of condensed atoms is equivalent to
a coherent state \cite{Yukalov_23}, these nonlinear excitations are called 
{\it nonlinear coherent modes}. These modes enjoy several important properties
allowing for the realization of interesting effects. A trapped Bose condensate with 
nonlinear coherent modes reminds an atom with several energy levels that can be 
connected by resonant electromagnetic field. Therefore a condensate with coherent 
modes allows for the occurrence of effects similar to those happening for atoms,
such as Rabi oscillations, appearance of interference patterns, Josephson current,
and Ramsey fringes. There also exist the effects of harmonic generation and 
parametric conversion.

At the same time, a Bose-condensed atomic system is nonlinear due to interactions 
between atoms, and it possesses highly nontrivial dynamics of coherent modes. It 
demonstrates different regimes of motion. In a Bose-condensed system, there is a 
mode-locked regime, when the modes are locked, so that their fractional populations 
vary only in a part of the whole admissible region. The other is mode-unlocked regime, 
when the mode populations vary in the whole admissible interval. The dynamic transition 
between these regimes is analogous to a phase transition in a statistical system. 
In a three-mode system, there can occurs the regime of chaotic motion, which is not 
possible in a two-mode system.   

For Bose-condensed system of trapped atoms, it is possible to realize collective 
squeezing effects. Simultaneous generation of coherent modes in several connected
traps, such as deep optical latices, allows for the creation of entangled mesoscopic
clouds and for the study of the effect of resonant entanglement production.

Coherent atomic clouds released from a trap form what is called atom laser. The 
difference of an atom laser from the photonic laser is that atoms fall down due to 
gravity, but cannot be forwarded in the desired direction. However, realizing special 
initial conditions one can trigger the beam emanation in any chosen direction. 

Nonequilibrium states can also be generated by strong nonresonant pumping creating
such coherent modes as quantum rings and vortices. Strong pumping leads to the
generation of vortex turbulence. Then there happens the formation of droplets of 
Bose condensate floating among uncondensed atoms, and droplet (or grain) turbulence 
develops. Even more strong pumping results in the destruction of condensate and wave
turbulence. The overall process of consecutive generation of nonequilibrium states
from an equilibrium condensate through ring and vortex states, to vortex and droplet
turbulence, and ending with uncondensed state, has been named {\it inverse 
Kibble-Zurek scenario}. A classification of nonequilibrium states is advanced, by 
means of relative injected energy, effective temperature, Fresnel and Mach numbers.          

Below, we use the system of units, where the Planck and Boltzmann constants are 
set to one.

\section{Condensate wave function}

Atomic systems Bose-condensed in traps are usually rather rarified, such that the 
effective interaction radius $r_{int}$ is much smaller than the mean interparticle 
distance $a$, which can be written as
\be
\label{1}
 \frac{r_{int}}{a} ~\ll ~1 \; , \qquad \rho^{1/3} r_{int} ~\ll ~1 \;  ,
\ee
where $\rho \sim a^{-3}$ is the average density of particles. At the same time the
effective strength of particle interactions 
\be
\label{2}
 \gm ~ \equiv ~\rho^{1/3}a_s  
\ee
can be large, with $a_s$ being scattering length. In that sense, the rarified system 
of trapped atoms is a gas cloud. 

The energy Hamiltonian has the standard form
\be
\label{3}
\hat H ~ = ~ 
\int \psi^\dgr(\br) \; \left ( -\; \frac{\nabla^2}{2m} + U \right) \; 
\psi(\br) \; d\br ~ + ~
\frac{1}{2} \int  \psi^\dgr(\br) \;  \psi^\dgr(\br') \; \Phi(\br-\br') \;
\psi(\br') \; \psi(\br) \; d\br d\br'\; ,
\ee
where $U = U({\bf r}, t)$ is an external trapping potential. Short-range interactions 
of rarified atoms can be represented by the effective interaction potential
\be
\label{4}
\Phi(\br) ~ = ~\Phi_0 \dlt(\br) \qquad 
\left( \Phi_0 ~\equiv ~4\pi \; \frac{a_s}{m} \right) \; .
\ee 
The Heisenberg equation of motion for the field operator reads as
\be
\label{5}
i \; \frac{\prt}{\prt t} \; \psi(\br,t) ~ = ~ 
H [\; \psi \; ] \; \psi(\br,t) \;  ,
\ee
with the effective nonlinear Hamiltonian
\be
\label{6}
H [\; \psi \; ] ~ \equiv ~ -\; \frac{\nabla^2}{2m} +  U(\br,t) +
\int \psi^\dgr(\br') \; \Phi(\br-\br') \; \psi(\br') \; d\br' \; .
\ee

Bose-Einstein condensation means the appearance of a coherent state. At 
asymptotically low temperature, such that 
\be
\label{7}
T ~ \ll ~ \rho \Phi_0 \; ,
\ee
and very weak interactions, when
\be
\label{8}
 \gm ~ \ll ~  1 \; ,
\ee
almost all atoms are Bose condensed, hence practically all the system is in a
coherent state. 

Coherent states are the eigenstates of the annihilation field operator $\psi(\br,t)$,
\be
\label{9}
\psi(\br,t) \; | \; \eta \; \rgl ~ = ~  \eta(\br,t) \; | \; \eta \; \rgl \; ,
\ee
with $\eta(\br,t)$ being called the coherent field or condensate wave function. 
The coherent field is the vacuum state of a Bose-condensed system \cite{Yukalov_23}. 
Averaging Eq. (\ref{5}) over the vacuum coherent state yields the equation for the 
condensate function (coherent field)
\be
\label{10}
i \; \frac{\prt}{\prt t} \; \eta(\br,t) ~ = ~ 
H [\; \eta \; ] \; \eta(\br,t) \; ,
\ee
where the nonlinear Hamiltonian is
\be
\label{11}
H [\; \eta \; ] ~ = ~  -\; \frac{\nabla^2}{2m} + U(\br,t) +
\int \Phi(\br-\br') \; | \; \eta(\br',t) \; |^2 d\br' \;   .   
\ee

This equation, for an arbitrary interaction potential, was advanced by Bogolubov 
in his book \cite{Bogolubov_24} that has been republished numerous times (see, e.g. 
\cite{Bogolubov_25,Bogolubov_26,Bogolubov_27}) and studied in Refs. 
\cite{Gross_28,Gross_29,Gross_30,Gross_31,Wu_32,Pitaevskii_33,Gross_34}. By its 
mathematical structure, this is a nonlinear Schr\"{o}dinger equation \cite{Malomed_35}. 
For trapped atoms with the interaction potential (\ref{4}), it reads as
\be
\label{12}
i \; \frac{\prt}{\prt t} \; \eta(\br,t) ~ = ~ 
\left[ \; -\; \frac{\nabla^2}{2m} + U(\br,t) + 
\Phi_0 \; |\; \eta(\br,t) \; |^2 \; \right] \; \eta(\br,t) \; .
\ee
The condensate function is normalized to the total number of atoms,
$$
 \int |\; \eta(\br,t) \; |^2 \;  d\br  ~ = ~  N \; .
$$

Recall that this equation assumes that all atoms are Bose-condensed and temperature
is zero, therefore it can provide a reasonable approximation only for very low 
temperatures and asymptotically weak interactions, when inequalities (\ref{7}) and 
(\ref{8}) are valid. 

Since the number of atoms $N$ is conserved, it is possible to make the replacement
\be
\label{13}
 \eta(\br,t) ~ = ~  \sqrt{N} \; \vp(\br,t) \; ,
\ee
so that the normalization condition becomes
$$
 \int |\; \vp(\br,t) \; |^2 \;  d\br  ~ = ~  1 \; .
$$

The external potential can be separated into a constant part of a trapping potential
and its temporal modulation,
\be
\label{14}
 U(\br,t) ~ = ~ U(\br) + V(\br,t) \;  .
\ee
Thus we come to the equation
\be
\label{15}
 i \; \frac{\prt}{\prt t} \; \vp(\br,t) ~ = ~ 
\left[ \; \hat H[ \; \vp\; ] + V(\br,t) \; \right] \; \vp(\br,t) \; ,
\ee
with a nonlinear Hamiltonian
\be
\label{16}
\hat H[ \; \vp\; ] ~ = ~  -\; \frac{\nabla^2}{2m} + U(\br) + 
N \Phi_0 \; |\; \vp(\br,t) \; |^2 \;   .
\ee
The Cauchy problem (\ref{15}), generally speaking, does not have a unique solution,
since it does not satisfy the conditions of the Cauchy-Kovalevskaya theorem 
\cite{Vinogradov_36}. To find uniquely defined solutions, one has to concretize the
sought class of functions \cite{Calogero_37,Faddeev_38}. We shall be looking for 
nonlinear solutions that can be treated as analytical continuation for the solutions 
of the linear Schr\"{o}dinger equation.

\section{Coherent modes}

Let us consider the stationary solutions to Eq. (15). Substituting in (\ref{12}) 
the expression
\be
\label{17}
\eta(\br,t) ~ = ~ \eta(\br) e^{-i E t}
\ee
gives the stationary eigenproblem
\be
\label{18}
 \hat H[ \; \vp_n \; ] \; \vp_n(\br) ~ = ~  E_n \; \vp_n(\br)
\ee
defining a spectrum $E_n$, labeled with a multi-index $n$, and eigenfunctions 
$\varphi_n({\bf r})$. The eigenfunctions $\varphi_n({\bf r})$ are called 
{\it coherent modes} \cite{Yukalov_22,Yukalov_39,Bagnato_40,Yukalov_41,Yukalov_42}. 

Coherent modes are not compulsorily orthogonal, so that the scalar product
$$
(\vp_m, \; \vp_n) ~ = ~ \int \vp_m^*(\br) \; \vp_n(\br) \; d\br \;  ,
$$
generally, is not a Kronecker delta. These modes do not form a resolution of unity.
However, the set $\{\varphi_n({\bf r})\}$ is total \cite{Klauder_43}, and the closed 
linear envelope over the set of the coherent modes
\be
\label{19}
\cH  ~ = ~ {\rm span}_n \{ \; \vp_n(\br) \; \}
\ee
composes a Hilbert space \cite{Yukalov_42}. Then the solution to Eq. (\ref{15})
can be represented as the expansion
\be
\label{20}
 \vp(\br,t) ~ = ~ \sum_n b_n(t) \; \vp_n(\br) \;  .
\ee
    
Approximate stationary solutions to Eq. (\ref{18}) can be found in different 
ways. At very small interaction parameters $\gamma$, it is possible to use simple 
perturbation theory \cite{Serov_44}. For arbitrary gas parameters $\gm$, optimized 
perturbation theory is applicable \cite{Yukalov_45} or self-similar approximation 
theory \cite{Yukalov_46}.  

To find the temporal behaviour of $\varphi({\bf r},t)$, we need to specify the 
explicit form of the modulation potential $V({\bf r},t)$. There are two ways of 
generating nonequilibrium states in a system of trapped atoms, a resonant and 
non-resonant pumping. 

(i) {\it Resonant generation}. One way is by employing an external field 
oscillating with a frequency $\omega$ that would be in resonance with a transition 
frequency
\be
\label{21}
\om_{21} ~ \equiv ~ E_2 - E_1
\ee
between the ground-state level, with the energy $E_1$, and an excited level, with 
the energy $E_2$, so that the detuning be small,
\be
\label{22}
 \left| \; \frac{\Dlt\om}{\om_{21} } \; \right| ~ \ll ~ 1 \; ,
\qquad \Dlt\om  ~ \equiv ~  \om - \om_{21} \;.
\ee
An oscillating field can be produced by modulating the potential trapping the atoms. 
More generally, it is possible to use several external fields with their frequencies
being in resonance with several transition frequencies between different modes. Under 
resonant generation, the most important is to preserve the resonance condition (\ref{22}),
while the modulation field can be rather small.    

(ii) {\it Non-resonant generation}. The other way is to use a nonresonant but 
sufficiently strong modulation field. Then not a single or several selected modes are
excited but a bunch of modes is generated. If the generation process lasts during the
time much longer then the oscillation time, then the higher modes decay into the most 
stable one that prevails.

\section{Resonant generation}

The modulation field can be chosen in the form
\be
\label{23}
V(\br,t) ~ = ~ V_1(\br) \cos(\om t) + V_2(\br) \sin(\om t) \; ,
\ee
with a frequency $\omega$ satisfying condition (\ref{22}). This situation is 
similar to the resonance excitation of atoms in optics \cite{Allen_47}, because 
of which it is called {\it atom optics} \cite{Yukalov_48,Yukalov_49}. 

The coefficient $b_n(t)$ in (\ref{20}) can be written as 
\be
\label{24}
 b_n(t)  ~ = ~  c_n(t) \exp( -i E_n t ) \; ,
\ee
where $c_n(t)$ is a slow function of time, as compared to the quickly oscillating 
exponential,
\be
\label{25}
 \frac{1}{E_n} \; \left| \; \frac{d c_n(t)}{dt} \; \right| ~ \ll ~ 1 \; .
\ee
This condition is analogous to the condition of slowly varying amplitude approximation 
in optics \cite{Allen_47,Mandel_50}.

Then we substitute presentation (\ref{24}) into equation (\ref{15}) and 
use the averaging techniques \cite{Bogolubov_51,Murdock_52} treating $c_n(t)$ 
as quasi-invariants \cite{Yukalov_53}. In this way, we need the notations for 
the transition amplitudes, due to atomic interactions,
\be
\label{26}
\al_{mn} ~ \equiv ~ \Phi_0 N \int |\; \vp_m(\br) \; |^2 \left[\; 
2\; |\; \vp_n(\br) \; |^2 - |\; \vp_m(\br) \; |^2 \; \right] \; d\br \; ,
\ee
and to the modulation field,
\be
\label{27}
\bt_{mn} ~ \equiv ~ \int \vp_m^*(\br) \left[ \; V_1(\br) - i V_2(\br) \; \right]
\; \vp_n(\br) \; d\br \;  .
\ee
Note that (\ref{27}) is an analog of the Rabi frequency.

In order that the averaging techniques be valid, one needs the inequalities
\be
\label{28}
\left| \; \frac{\al_{mn}}{\om_{mn} } \; \right| ~ \ll ~ 1 \; ,
\qquad
\left| \; \frac{\bt_{mn}}{\om_{mn} } \; \right| ~ \ll ~ 1 \;   ,
\ee
where $\omega_{mn} \equiv E_m - E_n$. Averaging over time, we use the equality
$$
\lim_{\tau\ra\infty} \; \frac{1}{\tau} 
\int_0^\tau \exp ( i\om_{mn} t) \; dt  ~ = ~ \dlt_{mn} \; .
$$
Thus wee come to the equations
\be
\label{29}
 i \; \frac{d c_n}{dt} ~ = ~ \sum_{m(\neq n)} \al_{mn} |\; c_m \; |^2 c_n
+ \frac{1}{2} \; \dlt_{n1} \bt_{12} c_2 e^{i\Dlt\om t} +
\frac{1}{2} \; \dlt_{n2} \bt_{12}^* c_1 e^{-i\Dlt\om t} \;  ,
\ee
with $c_n = c_n(t)$ satisfying the normalization condition
\be
\label{30}
 \sum_n |\; c_n(t) \; |^2 ~ = ~  1 \;  .
\ee

From these equations, we find that all levels, except $n=1$ and $n=2$ are not 
populated,
\be
\label{31}
c_n(t) ~ = ~ 0  \qquad ( n = 3,4,5,\ldots) \;  ,
\ee
while for the levels $n=1,2$, we have the equations 
$$
i\; \frac{d c_1}{dt} ~ = ~ \al_{12} |\; c_2\; |^2 c_1 + 
\frac{1}{2} \; \bt_{12} c_2 e^{i\Dlt\om t} \; ,
$$
\be
\label{32}
i\; \frac{d c_2}{dt} ~ = ~ \al_{21} |\; c_1\; |^2 c_2 + 
\frac{1}{2} \; \bt_{12}^* c_1 e^{- i\Dlt\om t} \;  .
\ee

Thus, it is shown that in the case of resonance generation, with one modulation 
field, only two modes are involved in dynamics. The properties of the coherent 
modes, in addition to the literature cited above, also have been studied theoretically 
in several articles 
\cite{Marzlin_54,Marzlin_55,Ostrovskaya_56,Feder_57,Kivshar_58,Damski_59,Agosta_60}. 
A nonlinear dipole mode in a two-component condensate was excited in experiment 
\cite{Williams_61}.

\section{Counterflow instability}

The appearance of coherent modes is due to strong fluctuations of density caused 
by an external alternating field. The density distribution inside the trap becomes 
essentially nonuniform. Different parts of the trapped atomic cloud quickly move 
with respect to each other, which leads to the occurrence of a kind of counterflow 
instability 
\cite{Yukalov_16,Nepomnyashchy_1974,Yukalov_1980,Yukalov_2004,Abad_2015,Yukalov_2022}
arising when different parts of a liquid system move with respect to each other.  
There exist several types of hydrodynamic instabilities occurring in moving liquids
\cite{Chandrasekhar_1961,Kul_1991}. For example, the Kelvin-Helmholtz instability 
occurs when there is a velocity difference across the interface between two fluids.
The Rayleigh-Taylor instability is an instability of an interface between two moving 
fluids of different densities which occurs when the lighter fluid is pushing the heavier 
fluid. Counterflow instabilities appear in moving Bose systems 
\cite{Yukalov_16,Nepomnyashchy_1974,Yukalov_1980,Yukalov_2004,Abad_2015,Yukalov_2022,
Sasaki_2009,Takeuchi_2010,Ishino_2011,Fujimoto_2012}.

The illustration of the appearance of counterflow instability in Bose-condensed systems
can be conveniently done by considering, first, a mixture of two or more Bose systems. 
In the local-density picture, the mixture components move with the corresponding velocities 
$v_i({\bf r,t})$. Collective excitations in the mixture can be found by the standard 
methods, either by linearizing the equations of motion or resorting to the random-phase 
approximation \cite{Yukalov_16,Yukalov_1980,Yukalov_2004,Yukalov_2022}. Assume that atoms 
interact through the local potentials
\be
\label{C1}  
\Phi_{ij}(\br) ~ = ~ \Phi_{ij} \dlt(\br) \; , \qquad 
\Phi_{ij} ~ \equiv ~ \int \Phi_{ij}(\br) \; d\br) \;   ,
\ee
in which
\be
\label{C2}
\Phi_{ij} ~ = ~ 4\pi \; \frac{a_{ij}}{m_{ij}} \; ,
\qquad
 m_{ij} ~ = ~  \frac{2m_im_j}{m_i+m_j} \; ,
\ee
where $m_i$ and $a_i$ are masses and scattering lengths of the related atoms.    

Inside a trap, the motion of different components can be organized by modulating the
trapping potential, e.g., according to the law
$$
U_i(\br,t) ~ = ~ \frac{1}{2} \; m_i \om_x^2 \; \left( x^2 + y^2 \right) +
 \lbd_i m_i^2 \om_z^3 \; \left[\; z^2 - b_i^2(t) \; \right] \; .
$$

In the long-wave limit, the collective spectra of each of the component, in the 
local-density approximation, have the acoustic form with the sound velocities defined
by the relation
\be
\label{C3}
 s_i^2 ~ = ~  \frac{\rho_j}{m_i} \; \Phi_{ii} \;   ,
\ee
where $s_i = s_i({\bf r},t)$ and $\rho_i = \rho_i({\bf r},t)$. The mixture of two
components is stable provided the stability condition is true 
\cite{Yukalov_1980,Yukalov_2004,Yukalov_2022}:
\be
\label{C4}
 s_{12}^2 ~ < ~ \left( s_1^2 - v_1^2 \right) \left( s_2^2 - v_2^2 \right) \; ,
\ee
where
\be
\label{C5}
  s_{12}^2 ~ = ~  s_{12}^2(\br,t) ~ = ~ 
\sqrt{ \frac{\rho_1(\br,t)\rho_2(\br,t)}{m_1m_2} } \; \Phi_{12} .
\ee
The stability condition can be rewritten as 
\be
\label{C6}
\frac{\Phi_{12}^2}{\Phi_{11}\Phi_{22}} ~ < ~ 
\left( 1 - \; \frac{m_1 v_1^2}{\rho_1\Phi_{11}} \right) 
\left( 1 - \; \frac{m_2 v_2^2}{\rho_2\Phi_{22}} \right) \;   .
\ee

Instead of considering a mixture of different components, it is admissible to treat 
a nonequilibrium system as a mixture of different coherent modes, with $m_i = m_j \equiv m$.
In each mode, the atomic interactions are the same, $\Phi_{ii} = \Phi_{jj} \equiv \Phi_0$. 
In order to distinguish the modes, one needs to assume that at each spatial point, at each
given time, there can exist a single mode, so that 
$\rho_i({\bf r},t) \rho_j({\bf r},t) = 0$ for $i \neq j$. Then $s_{ij} = 0$ if $i \neq j$.
As a result, one gets the stability condition
\be
\label{C7}
 0 ~ < ~ \left( s_1^2 - v_1^2 \right) \left( s_2^2 - v_2^2 \right) \;  .
\ee
In that way, the system stability becomes lost as soon as one of the velocities exceeds the
related sound velocity.

\section{Energy levels}

The mode energy is defined by the eigenproblem (\ref{18}). The energy levels can be 
found using optimized perturbation theory (see reviews \cite{Yukalov_62,Yukalov_63}). 
For instance, let us consider a cylindrical trap with the radial frequency
$\omega_x = \omega_y$ and the axial frequency $\omega_z$. The related aspect ratio 
is denoted as
\be
\label{33}
\nu ~ \equiv ~ \frac{\om_z}{\om_x} ~ = ~ \frac{l_x^2}{l_z^2} \; .
\ee
It is convenient to use the dimensionless cylindrical variables
\be
\label{34}
  r ~ \equiv ~ \frac{\sqrt{r_x^2+r_y^2}}{l_x} \; , 
\qquad
z ~ \equiv ~ \frac{r_z}{l_x} 
\qquad 
\left ( l_x \; \equiv \; \frac{1}{\sqrt{m\om_x} }  \right) 
\ee
and the dimensionless coupling parameter
\be
\label{35}
g ~ \equiv ~ 4\pi \; \frac{a_s}{l_x} \; N \; .
\ee

The nonlinear Hamiltonian (\ref{16}), measured in units of the radial frequency, is
\be
\label{36}
 \hat H ~ \equiv ~ \frac{\hat H[\;\vp\;]}{\om_x} ~ = ~ - \; 
\frac{\nabla^2}{2} + \frac{1}{2} \; \left( r^2 + \nu^2 z^2 \right)
+ g |\; \psi \;|^2 \;  ,
\ee
where the dimensionless wave function is
\be
\label{37}
\psi(r,\vp,z) ~ \equiv ~ l_x^{3/2} \vp(\br) \; .
\ee
In dimensionless units, the eigenproblem (\ref{18}) reads as
\be
\label{38}
 \hat H \psi_{nmj}(r,\vp,z) ~ = ~ E_{nmj} \psi_{nmj}(r,\vp,z) \; ,
\ee
where $n=0,1,2, \ldots$ is the radial quantum number, $m=0,\pm 1,\pm 2,\ldots$ 
is the azimuthal quantum number, and $j=0,1,2, \ldots$ is the axial quantum 
number.

Then we use the Rayleigh-Schr\"{o}dinger perturbation theory, starting with the 
initial Hamiltonian
\be
\label{39}
 \hat H_0  ~ = ~ -\;\frac{\nabla^2}{2} +  
\frac{1}{2} \; \left( u^2 r^2 + v^2 z^2 \right)
\ee
and calculating the spectrum in the orders $k=1,2,\ldots$,
$$
E_{nmj}^{(k)} ~ = ~ E_{nmj}^{(k)}(g,u,v) \;  .
$$
The zero-order wavefunctions are
$$
\psi_{nmj}^{(0)}(r,\vp,z) ~ = ~ \left[\; 
\frac{2n! u^{m+1}}{(n + |\; m\;| )!} \; \right]^{1/2} \; r^{|m|} \;
\exp\left( -\; \frac{u}{2} \; r^2 \right)\; \times
$$
\be
\label{40}
\times \;
L_n^{|m|}(ur^2) \; \frac{e^{im\vp}}{\sqrt{2\pi}} \;
\left( \frac{v}{\pi} \right)^{1/4} \; 
\frac{1}{\sqrt{2^j j!} } \; \exp\left( -\; \frac{v}{2} \; z^2 \right)\;
H_j(\sqrt{v}\; z ) \;   ,
\ee
where $L_m^n$ is a Laguerre polynomial and $H_j$ is a Hermit polynomial. Here 
$u$ and $v$ are control parameters defined so that to control the convergence 
of perturbation theory.

Calculating the energy $E_{nmj}^{(k)}$ in a $k$-th order of perturbation theory, 
we define the control functions $u = u_k(g)$ and $v = v_k(g)$  by the conditions 
\be
\label{41}
\frac{\prt}{\prt u} \; E_{nmj}^{(k)} ~ = ~ 0 \; , \qquad
\frac{\prt}{\prt v} \; E_{nmj}^{(k)} ~ = ~ 0 \;   .
\ee
We introduce the effective coupling 
\be
\label{42}
 \widetilde g ~ \equiv ~ 2 ( 2 n + |\;m\; | + 1 ) \; \sqrt{2j + 1}\;
I_{nmj} g \nu \; ,
\ee
in which
$$
I_{nmj} ~ \equiv ~ \frac{1}{u\sqrt{v} } \int
|\; \psi_{nmj}(r,\vp,z) \; |^4 \; r \; dr d\vp dz \;  .
$$
 
For illustration, we may give the behavior of the energy spectrum obtained in 
first-order optimized perturbation theory. Thus for weak coupling, we have
\be
\label{43}
E_{nmj} ~ \simeq ~ a_0 + a_1 \widetilde g \qquad ( \widetilde g \ra 0) \;   ,
\ee
where
$$
a_0 ~ = ~ ( 2 n + |\; m\;| + 1) + \frac{2j+1}{2} \; \nu \; ,
\qquad
a_1 ~ = ~  \frac{1}{2( 2n + |\; m\;| + 1)\; \sqrt{(2j+1)\nu} } \; ,
$$
and for strong coupling, we get
\be
\label{44}
 E_{nmj} ~ \simeq ~ b_0 \widetilde g^{2/5} + b_1 \widetilde g^{-2/5} 
\qquad ( \widetilde g \ra \infty ) \; ,
\ee
where
$$
b_0 ~ = ~ \frac{5}{4} \; , \qquad
b_1 ~ = ~ \frac{1}{2} \; ( 2n + |\; m\;| + 1)^2 +
\frac{1}{4} \; ( 2j + 1)^2 \; \nu^2 \;   .
$$
Details can be found in \cite{Yukalov_42}. The typical situation for trapped 
atoms is the strong-coupling regime.

\section{Mode dynamics}

For studying the mode dynamics, it is convenient to introduce the population 
difference
\be
\label{45}
s ~ \equiv ~ |\; c_2\; |^2 - |\; c_1\; |^2
\ee
varying in the interval $-1 \leq s \leq 1$. Then the mode populations can be written as
$$
 |\; c_1\; |^2 ~ = ~ \frac{1-s}{2} \; , \qquad 
 |\; c_2\; |^2 ~ = ~ \frac{1+s}{2} \;  ,
$$
which defines the presentation
\be
\label{46}
 c_1 ~ = ~ \sqrt{ \frac{1-s}{2} } \; 
\exp\left\{ i \left( \zeta_1 + \frac{\Dlt\om}{2} \; t \right) \right\} \; ,
\qquad
c_2 ~ = ~ \sqrt{ \frac{1+s}{2} } \; 
\exp\left\{ i \left( \zeta_2 - \; \frac{\Dlt\om}{2} \; t \right) \right\} \; ,
\ee
where $\zeta_i = \zeta_i(t)$ are real-valued phases. 

Also, we introduce the notations for the average interaction amplitude
\be
\label{47}
\al ~ \equiv ~ \frac{1}{2} \; ( \al_{12} + \al_{21} ) \;  ,
\ee
the transition amplitude
\be
\label{48}
\bt_{12} ~ = ~ \bt e^{i\gm} \; , \qquad \bt ~ \equiv ~ |\; \bt_{12}\; | \;   ,
\ee
effective detuning
\be
\label{49}
 \dlt~ \equiv ~ \Dlt\om + \frac{1}{2} \;  ( \al_{12} - \al_{21} ) \;  ,
\ee
and the phase difference
\be
\label{50}
x ~ \equiv ~ \zeta_2 - \zeta_1 + \gm \; .
\ee

In terms of the above notations, equations (\ref{32}) can be represented as
$$
\frac{ds}{dt} ~ = ~ - \bt \; \sqrt{1-s^2} \; \sin x \; ,
$$
\be
\label{51}
 \frac{dx}{dt} ~ = ~ \al s + 
\frac{\bt s}{ \sqrt{1-s^2} }  \; \cos x + \dlt \;  .
\ee
These equations can be rewritten in the Hamiltonian form
\be
\label{52}
\frac{ds}{dt} ~ = ~ - \; \frac{\prt H}{\prt x} \; , 
\qquad
\frac{dx}{dt} ~ = ~  \frac{\prt H}{\prt s} \;   ,
\ee
with the Hamiltonian
\be
\label{53}
 H ~ = ~ \frac{\al}{2} \; s^2 \; - \; 
\bt \; \sqrt{1-s^2} \; \cos x + \dlt s \;  .
\ee
The population difference  varies in the interval $-1 \leq s \leq 1$ and the phase,
in the interval $0 \leq x \leq 2\pi$. 
  
We shall need the notations for the dimensionless Rabi frequency and detuning
\be
\label{54}
b ~ \equiv ~ \frac{\bt}{\al} \; , \qquad  
\ep ~ \equiv ~ \frac{\dlt}{\al} \;  .
\ee
The dimensionless detuning, by choosing $\Delta \omega$, can always be made small, 
so that $\varepsilon \ll 1$.  

In the case of $|b| > 1$, Eqs. (\ref{51}) possess three fixed points
$$
s_1^* ~ = ~ \frac{\ep}{b} \; , \qquad x_1^* = 0 \; ,
$$
$$
s_2^* ~ = ~ -\; \frac{\ep}{b} \; , \qquad x_2^* = \pi \; ,
$$
\be
\label{55}
s_3^* ~ = ~ \frac{\ep}{b} \; , \qquad x_3^* = 2\pi \qquad 
( |\; b\; | ~ \geq ~ 1 ) \; .
\ee
The stability analysis shows that all these points are the centers.

When $0 \leq b < 1$, there are five fixed points, the three fixed points (\ref{55})
plus the points
$$
s_4^* ~ = ~ \sqrt{1-b^2} + \frac{b^2\ep}{1-b^2} \; , \qquad x_4^* = \pi \; ,
$$
\be
\label{56}
s_5^* ~ = ~ -\; \sqrt{1-b^2} + \frac{b^2\ep}{1-b^2} \; , \qquad x_5^* = \pi \; 
\qquad ( 0 ~ \leq b ~ < ~ 1)  .
\ee
The point $(s_2^*, x_2^*)$ is a saddle, and all other points are centers. 

For $-1 < b \leq 0$, there occur seven fixed points, the three points, as in (\ref{55}),
plus
$$
s_4^* ~ = ~ \sqrt{1-b^2} + \frac{b^2\ep}{1-b^2} \; , \qquad x_4^* = 0 \; ,
$$
$$
s_5^* ~ = ~ -\; \sqrt{1-b^2} + \frac{b^2\ep}{1-b^2} \; , \qquad x_5^* = 0 \; ,
$$
$$
s_6^* ~ = ~ s_4^* \; , \qquad x_6^* ~ = ~ 2\pi \; ,
$$
\be
\label{57}
 s_7^* ~ = ~ s_5^* \; , \qquad x_7^* ~ = ~ 2\pi \qquad ( - 1 ~ < ~ b ~ \leq 0 ) \;  .
\ee
The point $(s_2^*, x_2^*)$ remains a saddle, while all others are the centers. 

The trajectory, passing through a saddle point, is called the {\it saddle separatrix}.
This line separates the phase plane onto the regions of qualitatively different 
regimes of motion. The separatrix equation is
\be
\label{58}
 \frac{s^2}{2} \; - \; b \; \sqrt{1-s^2} \; \cos x + \ep s ~ = ~ b \;  .
\ee
Here and in the equations above, the smallness of $\varepsilon \ll 1$ is used.
More details on the mode dynamics are given in Refs.
\cite{Yukalov_22,Yukalov_42,Yukalov_48,Yukalov_49}. The resonant excitation of the 
chosen mode can continue during the time of order $\omega_{21}/\beta^2$, after which
the effect of power broadening becomes essential, when neighboring modes become 
involved in the process.

\section{Mode locking}

When the separatrix crosses the point of initial conditions $\{s_0,x_0\}$, the 
trajectory moves from one region of the phase space to another region, which implies 
an abrupt qualitative change of dynamics. The separatrix, crossing the point of 
initial conditions, defines the {\it critical line} corresponding to the critical 
values of the parameters
\be
\label{59}
 \frac{s_0^2}{2} \; - \; 
b_c \; \sqrt{1-s_0^2} \; \cos x_0 + \ep_c s_0 ~ = ~ b_c \; .
\ee
Without the loss of generality, the initial condition for the phase can be taken 
as $x_0 = 0$. Under the fixed initial conditions and detuning $\varepsilon_c$, the 
critical pumping field reads as
\be
\label{60}
b_c ~ = ~ \frac{s_0^2+2\ep_c s_0}{2(1+\sqrt{1-s_0^2})} \;   .
\ee
Crossing the critical line (\ref{59}), the regime change happens abruptly, when $b$ 
moves between the regions with $b < b_c$ and $b > b_c$. When $b < b_c$, the effect 
of {\it mode locking} exists, such that the variation of the population difference 
is locked in a half of the allowed interval,
$$
- 1 ~ \leq ~ s ~ \leq ~ 0 \qquad ( b < b_c, \; s_0 = -1 ) \; ,
$$
\be
\label{61}
0 ~ \leq ~ s ~ \leq ~ 1 \qquad ( b < b_c, \; s_0 = 1 ) \;   ,
\ee
depending on initial conditions. And if $b > b_c$ then the period of the population
difference oscillations doubles, and the interval of variation covers all allowed region,
\be
\label{62}
- 1 ~ \leq ~ s ~ \leq ~ 1 \qquad ( b > b_c ) \;   .
\ee
Thus the value $b_c$ corresponds to a dynamic phase transition point. If the initial 
condition for the population difference is $s_0 = - 1$, then the critical point
becomes
\be
\label{63}
 b_c ~ = ~ \frac{1}{2} \; - \; \ep_c \;  .
\ee 

The dynamic transition between mode-locked regime and mode-unlocked regime is 
illustrated in Fig. 1a (mode-locked regime), Fig. 1b (close to transition), and
Fig. 1c (mode-unlocked regime). Regular oscillations are presented in Fig. 1d.

\begin{figure}[ht]
\centerline{
\hbox{ \includegraphics[width=7.5cm]{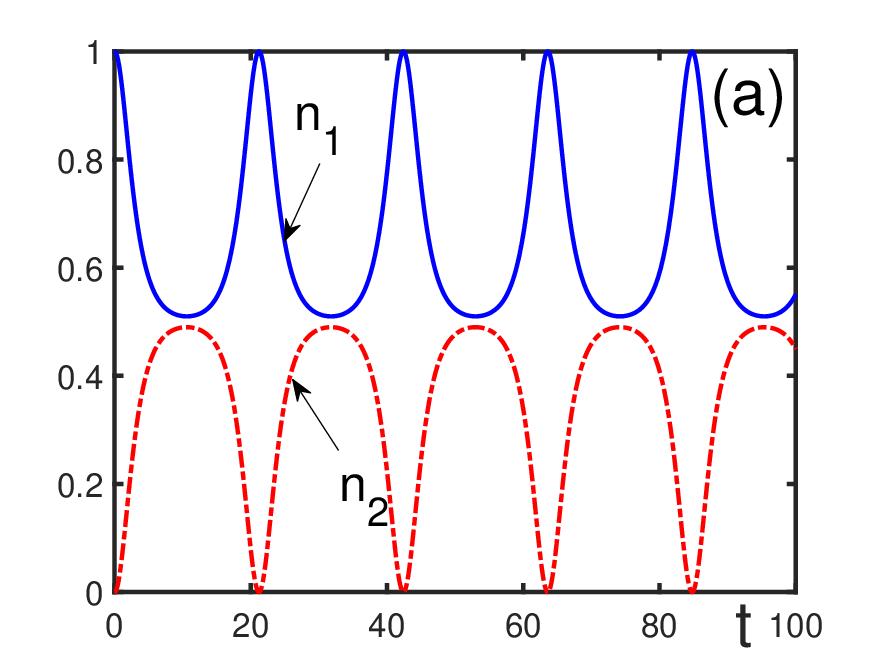} \hspace{1cm}
\includegraphics[width=7.5cm]{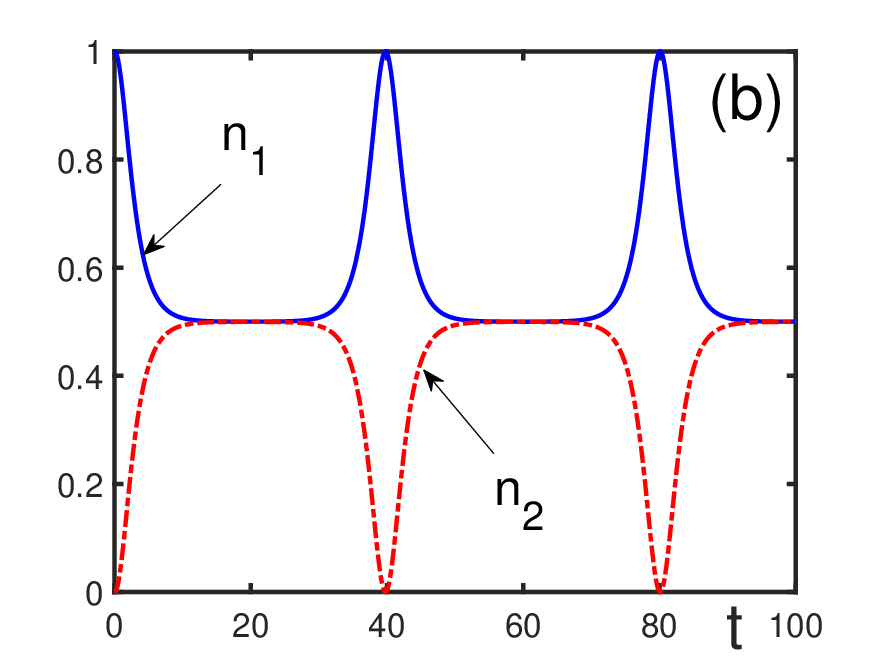}  } }
\vspace{12pt}
\centerline{
\hbox{ \includegraphics[width=7.5cm]{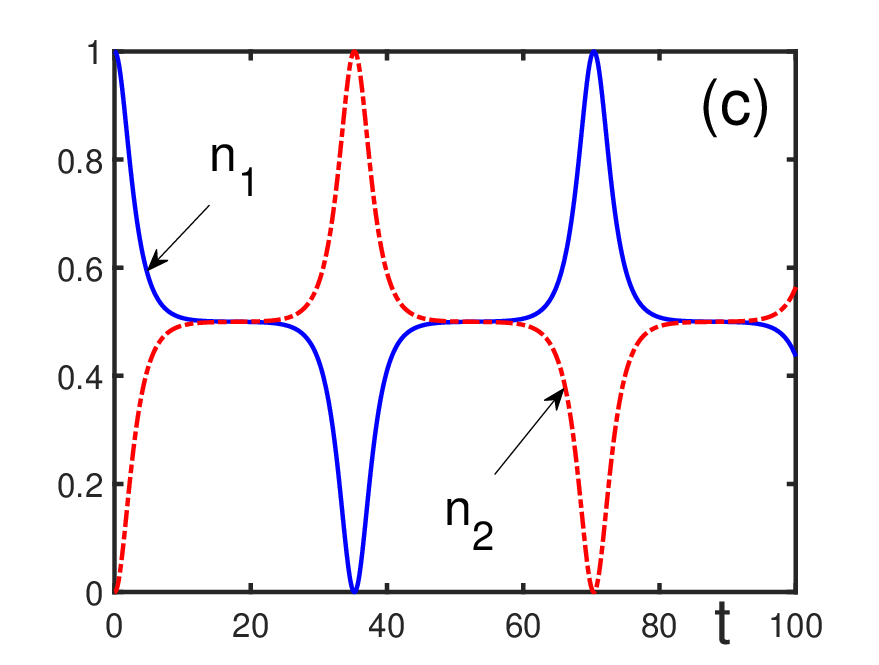} \hspace{1cm}
\includegraphics[width=7.5cm]{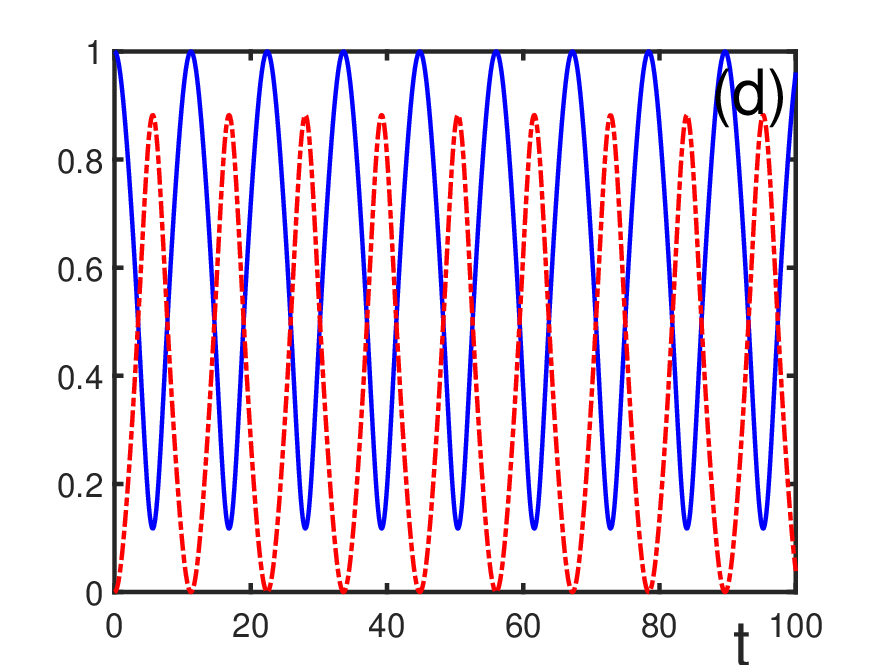} } }
\caption{\small
The time dependence of the fractional populations $n_1(t)$
(solid line) and $n_2(t)$ (dashed-dotted line) for $b=0.4999$ and
(a) $\dlt=0$;
(b) flattening of the fractional populations with oscillation period being
doubled for $\dlt=0.0001$;
(c) appearance of the upward cusps of $n_1(t)$ and downward cusps of $n_2(t)$ 
for $\dlt=0.0001001$;
(d) Regular oscillations for $\dlt=0.3$.
}
\label{fig:Fig.1}
\end{figure}

The qualitative change of the phase portrait is called {\it dynamical phase transition}.
In the vicinity of the dynamic critical line, there appear critical phenomena that are 
analogous to thermodynamic phase transitions. This can be shown by averaging the 
considered dynamical system over time \cite{Yukalov_41,Yukalov_64,Yukalov_65}.

The possibility of preserving resonance conditions is limited in time by a value
of order
\be
\label{64}
t_{res} ~ \sim ~ \frac{\om_{21}}{\al^2+|\;\bt_{12}\;|^2}    
\ee
that can be called {\it resonance time} \cite{Yukalov_42}. After this time, the effect 
of power broadening comes into play, when other energy levels become involved, so that 
good resonance cannot be anymore supported. This limiting resonance time, in realistic 
trapping experiments, is of order of $10$ seconds, which is comparable to the lifetime 
of atoms in a trap \cite{Courteille_6,Barrett_66,Ryu_67}.

\section{Shape-conservation criterion}

In order to transfer trapped atoms into an excited coherent mode, it is necessary to
apply an alternating field. However not any, even resonant, field can generate a
coherent mode. Some types of fields, would merely move the trapped atomic cloud without
generating excited modes. The situation when the cloud does not change its shape,
hence when no coherent modes are generated, despite the action of an alternating field,
is described by the following theorem \cite{Yukalov_68}.

\vskip 2mm

{\bf Theorem}. {\it Let atoms be trapped in a potential $U({\bf r})$, so that their 
state satisfies the trapping condition
\be
\label{65}
 \lim_{r\ra\infty} \; \vp(\br,t) ~ = ~ 0 \;  .
\ee
And let these atoms, being at the initial moment of time in a real state 
$\varphi_0({\bf r}) = \varphi_0^*({\bf r})$, be subjected to the action of an 
alternating field $V({\bf r},t)$. Then the solution to the nonlinear Schr\"{o}dinger 
equation (\ref{15}) satisfies the shape-conservation condition   
\be
\label{66}
 |\; \vp(\br,t) \; | ~ = ~  |\; \vp(\br-\ba(t),0) \; | \;  ,
\ee 
where $a(t)$ is a function of time, if and only if the trapping potential is harmonic,
\be
\label{67}
U(\br) ~ = ~ A_0 + {\bf A}_1 \br + \sum_{\al\bt} A_{\al\bt}\; r^\al\; r^\bt \; ,
\ee
and the alternating field is linear with respect to the spatial variables,
\be
\label{68}
 V(\br,t) = B_0(t) + {\bf B}_1(t) \cdot \br \; ,
\ee
with $B_0(t)$ and ${\bf B}_1(t)$ being arbitrary functions of time.}  

\vskip 2mm

In order to efficiently excite nonlinear coherent modes, it is required that at least
one of the conditions (\ref{65}) to (\ref{68}) be invalid.

\section{Multiple mode generation}

It is feasible to generate not just a single coherent mode, but several of them. For
this purpose, one needs to invoke an alternating field with several frequencies that
are in resonance with the related transition frequencies. Generally, a multimode
alternating field can be written in the form
\be
\label{69}
 V(\br,t) ~ = ~ \frac{1}{2} \sum_j \left[ \; B_j(\br) e^{i\om_j t} + 
B_j^*(\br) e^{-i\om_j t} \; \right] \;  ,
\ee 
with the modulation frequencies $\omega_j$ tuned to some transition frequencies 
$\omega_{mn}$. For example, the excitation of two coherent modes can be realized using 
one of three schemes that, similar to the optical excitation schemes \cite{Mandel_50}, 
can be called cascade, $V$-type and $\Lambda$-type.

In the {\it cascade scheme}, one couples the first (ground-state) level with the 
second energy level and the second, with the third level by employing the 
alternating-field frequencies in resonance with the related energy differences:
\be
\label{70}
\om_1 ~ = ~ \om_{21} \; , \qquad  \om_2 ~ = ~ \om_{32} \qquad (cascade)  .
\ee
Here exact resonance conditions are defined. Generally, there can exist small detunings
from the resonance.
   
In the {\it $V$-type scheme}, two alternating-field frequencies induce resonance with 
the transition frequencies, so that
\be
\label{71}
\om_1 ~ = ~ \om_{21} \; , \qquad  \om_2 ~ = ~ \om_{31} \qquad (V - type)    .
\ee

And in the {\it $\Lambda$-type scheme}, the coupling of the energy levels is realized
by means of the resonance
\be
\label{72}
\om_1 ~ = ~ \om_{31} \; , \qquad  \om_2 ~ = ~ \om_{32} \qquad (\Lambda - type)   .
\ee
 
As in the case of two coherent modes, the solution to equation (\ref{15}) is represented
as the expansion
\be
\label{73}
 \vp(\br,t) ~ = ~ \sum_n c_n(t) \; \vp_n(\br) \; e^{-iE_n t} \;  .
\ee
Then, following the same way as for two modes, we obtain the equations
$$
i\; \frac{dc_1(t)}{dt} ~ = ~ 
\left( \al_{12} |\; c_2\; |^2 +  \al_{13} |\; c_3\; |^2 \right)\; c_1 + f_1 \; ,
$$
$$
i\; \frac{dc_2(t)}{dt} ~ = ~ 
\left( \al_{21} |\; c_1\; |^2 +  \al_{23} |\; c_3\; |^2 \right) \; c_2 + f_2 \; ,
$$
\be
\label{74}
 i\; \frac{dc_3(t)}{dt} ~ = ~ 
\left( \al_{31} |\; c_1\; |^2 +  \al_{32} |\; c_2\; |^2 \right) \; c_3 + f_3 \;  ,
\ee
in which the functions $f_j$ depend on the type of the generation scheme. The related
detunings $\Delta \omega_{mn}$ from the resonance are assumed to be small
\be
\label{75}
 \left| \; \frac{\Dlt\om_{mn}}{\om_{mn}} \; \right| ~ \ll ~ 1 \; , \qquad
( \Dlt\om_{mn} \equiv \om_j - \om_{mn} ) \;  .
\ee
For an example of the functions $f_j$, let us write down them for the cascade 
generation:
$$
f_1 ~ = ~ \frac{1}{2} \; \bt_{12} \; c_2 \; e^{i\Dlt\om_{21}t} \; ,
$$
$$
f_2 ~ = ~ \frac{1}{2} \; \bt_{12}^* \; c_1 \; e^{-i\Dlt\om_{21}t} +
\frac{1}{2} \; \bt_{23} \; c_3 \; e^{i\Dlt\om_{32}t} \; ,
$$
\be
\label{76}
f_3 ~ = ~ \frac{1}{2} \; \bt_{23}^* \; c_2 \; e^{-i\Dlt\om_{32}t}   ,
\ee
where
$$
 \Dlt\om_{21} ~ \equiv ~ \om_1 - \om_{21} \; , \qquad 
 \Dlt\om_{32} ~ \equiv ~ \om_2 - \om_{32} \; .
$$

The number of modes that could be excited is defined by the available number of 
states bound inside the trapping potential. For an infinitely rising potential, there
is infinite number of state. Of course, in real experiments, trapping potentials are
finite, hence they possess a finite number of states \cite{Simon_69}. 

Solutions in the case of three modes have been analyzed in Refs. 
\cite{Yukalov_68,Yukalov_70,Yukalov_71} using two ways, resorting to the 
averaging techniques \cite{Bogolubov_51} and by direct numerical solution of the 
nonlinear Schr\"{o}dinger equation (\ref{15}). Both ways were found to agree well 
with each other. 

In the case of two coexisting nonlinear modes, the system of two complex 
differential equations (\ref{32}), taking account of the normalization condition 
(\ref{30}), has been reduced to two real differential equations (\ref{51}). 
A two-dimensional dynamical system cannot exhibit chaotic motion. In our case, the
fixed points where either centers or saddles.

For the case of three nonlinear modes, three complex equations (\ref{74}) can be 
reduced to the system of four real differential equations. For this purpose, we 
introduce the real-valued phases $\zeta_n(t)$ by the representation
\be
\label{77}
c_n ~ \equiv ~ |\; c_n \; | \; \exp( i \zeta_n) \; .
\ee
We also use the notation
\be
\label{78}
 \bt_{mn} ~ = ~ b_{mn}\; \exp(i\gm_{mn}) \; , \qquad 
b_{mn} ~ \equiv ~ |\; \bt_{mn}\;| \;  .
\ee
Introduce the population differences 
\be
\label{79}
s ~ \equiv ~ |\; c_2\;|^2 - |\; c_1\;|^2 \; , \qquad
p ~ \equiv ~ |\; c_3\;|^2 - |\; c_2\;|^2 \; , 
\ee
and the relative phases
\be
\label{80}
 x ~ \equiv ~ \zeta_2 - \zeta_1 + \gm_{12} + \Dlt\om_{21} t \; , 
\qquad  
y ~ \equiv ~ \zeta_3 - \zeta_2 + \gm_{23} + \Dlt\om_{32} t \; .
\ee
Then the mode populations can be written as
$$
|\; c_1\;|^2 ~ = ~ \frac{1}{3} \; ( 1 - 2s - p ) \; , 
$$
\be
\label{81}
|\; c_2\;|^2 ~ = ~ \frac{1}{3} \; ( 1 + s - p ) \; , \qquad
|\; c_3\;|^2 ~ = ~ \frac{1}{3} \; ( 1 + s + 2p ) \;  .
\ee

Using these expressions makes it possible to reduce the system of six real equations 
(\ref{74}) to the system of four equations. For instance, in the case of the cascade 
generation, we get
$$
\frac{ds}{dt} ~ = ~ \frac{1}{3} \; \sqrt{1+s-p} \; (\; b_{23}\; \sqrt{1+s+2p} \; \sin y -
2 b_{12}\; \sqrt{1-2s-p} \; \sin x \; ) \; ,
$$
$$
\frac{dp}{dt} ~ = ~ \frac{1}{3} \; \sqrt{1+s-p} \; ( \; b_{12}\; \sqrt{1-2s-p} \; \sin x -
2 b_{23}\; \sqrt{1+s+2p} \; \sin y \; ) \; ,
$$
$$
\frac{dx}{dt} ~ = ~ \al_1\; s + \dlt_1 \; p + 
\frac{3b_{12} s\cos x}{2\sqrt{(1+s-p)(1-2s-p)} }
\; - \; \frac{1}{2} \; b_{23} \; \sqrt{\frac{1+s+2p}{1+s-p} } \; \cos y + \dlt_2 \; ,
$$
\be
\label{82}
\frac{dy}{dt} ~ = ~ \al_2\; p + \dlt_3 \; s + 
\frac{3b_{23} p\cos y}{2\sqrt{(1+s-p)(1+s+2p)} }
+ \frac{1}{2} \; b_{12} \; \sqrt{\frac{1-2s-p}{1+s-p} } \; \cos x + \dlt_4 \;  ,
\ee
where
$$
\al_1 ~ \equiv ~ \frac{1}{3} \; ( \al_{12} + \al_{13} + 2\al_{21} - \al_{23} ) \; ,
\qquad
\al_2 ~ \equiv ~ \frac{1}{3} \; ( \al_{32} + \al_{31} + 2\al_{23} - \al_{21} ) \; ,
$$
$$
\dlt_1 ~ \equiv ~ \frac{1}{3} \; ( \al_{21} - \al_{12} + 2\al_{13} - 2\al_{23} ) \; ,
\qquad
\dlt_2 ~ \equiv ~ \Dlt\om_{21} + 
\frac{1}{3} \; ( \al_{12} - \al_{21} + \al_{13} - \al_{23} ) \; ,
$$
$$
\dlt_3 ~ \equiv ~ \frac{1}{3} \; ( \al_{23} - \al_{32} + 2\al_{31} - 2\al_{21} ) \; ,
\qquad
\dlt_4 ~ \equiv ~ \Dlt\om_{32} + 
\frac{1}{3} \; ( \al_{23} - \al_{32} + \al_{21} - \al_{31} ) \;   .
$$
The four-dimensional dynamical system (\ref{82}) demonstrates the effects similar to 
those of the two-dimensional system (\ref{51}), such as {\it mode locking} for small
pumping amplitudes $b_{ij}$, however with quasi-periodic motion, instead of periodic 
Rabi oscillations. For large pumping amplitudes $b_{ij}$, there appears {\it chaotic 
motion}. For example, in the case where $\alpha_{mn} = \alpha$, $\beta_{mn} = \beta$,
and $\Delta \omega_{mn} = 0$, the chaotic motion arises when
$$
 \left| \; \frac{\bt}{\al} \; \right| ~ \geq ~ 0.639448 \;  .
$$
 
Note that here we consider cold bosonic atoms trapped in a potential that is not 
periodic, although similar effects can be generated in nonequilibrium optical latices 
as well \cite{Jauregui_72,Kumakura_73,Poli_74,Marzlin_75,Alberti_76,Parker_77,Zheng_78,
Zhang_79,Niu_80,Hudomal_81,Boulier_82,Zheng_83}.

\section{Matter-wave interferometry}

Composing a coherent system, Bose condensates enjoy many of the properties typical of 
coherent optical systems and studied in atom interferometry \cite{Cronin_84}. In atoms, 
electrons can be excited into a higher individual energy level, while in a condensate 
a large group of trapped Bose-condensed atoms can be transferred into an excited 
collective energy level. The principal difference in dealing with many atoms, instead 
of a single atom, is in the necessity of taking account of interactions of condensed 
atoms.

\subsection{Interference patterns}

The existence of different coherent modes inside a trap leads to the appearance of 
interference patterns \cite{Yukalov_42}. The density of atoms in a trap is
\be
\label{83}
\rho(\br,t) ~ = ~ |\; \eta(\br,t) \; |^2 ~ = ~ N \; |\; \vp(\br,t)\; |^2 \; .
\ee
With expression (\ref{20}), this gives
\be
\label{84}
\rho(\br,t) ~ = ~ N 
\left| \; \sum_n c_n(t) \; e^{-i E_n t} \; \vp_n(\br) \; \right|^2 \;   .
\ee
The density of an $n$-th coherent mode is
\be
\label{85}
 \rho_n(\br,t) ~ = ~ N \; |\; c_n(t) \; \vp_n(\br)\; |^2 \;   .
\ee
Therefore the interference pattern 
\be
\label{86}
 \rho_{int}(\br,t) ~ \equiv ~  \rho(\br,t) - \sum_n  \rho_n(\br,t)
\ee
reads as
\be
\label{87}
\rho_{int}(\br,t) ~ = ~ N 
\sum_{m\neq n} c_m^*(t) \; c_n(t) \; \vp_m^*(\br) \; \vp_n(\br) \; e^{i\om_{mn}t}    .
\ee
Since the density of atoms can be observed, the interference pattern is also observable.

\subsection{Internal Josephson current}

The appearance of several coherent modes inside a trap makes the trapped atomic cloud
essentially nonuniform, since different modes have different topological shapes. 
Respectively, inside the trap there arise a density current that is analogous to the 
current occurring between different potential wells \cite{Milburn_85,Raghavan_86}.
The current appearing not because of external barriers, but as a consequence of 
internally arising nonuniformities, is called {\it internal Josephson current}
\cite{Legget_87,Ohberg_88}. This effect is also named {\it quantum dynamical 
tunneling} \cite{Davis_89,Heller_90}.

The density current in a condensed system reads as
\be
\label{88}
 {\bf j}(\br,t) ~ \equiv ~ 
\frac{1}{m} \; {\rm Im} \; \eta^*(\br,t) \; \vec{\nabla}\eta(\br,t) ~ = ~ 
\frac{N}{m} \; {\rm Im} \; \vp^*(\br,t) \; \vec{\nabla}\vp(\br,t) \; .
\ee
The current caused by a single mode is
\be
\label{89}
 {\bf j}_n(\br,t) ~ = ~  \frac{N}{m} \; {\rm Im} \; |\; c_n(t)\; |^2 \; 
\vp_n^*(\br,t) \; \vec{\nabla}\vp_n(\br,t) \;  .
\ee
The difference between the total current in the system and the sum of density currents
of separate modes forms the internal Josephson current \cite{Yukalov_42}
\be
\label{90}
 {\bf j}_{int}(\br,t) ~ = ~  \frac{N}{m} \; {\rm Im} \sum_{m\neq n}
c_m^*(t) \; c_n(t) \; e^{i\om_{mn} t} \;
\vp_m(\br,t) \; \vec{\nabla}\vp_n(\br,t) \;   .
\ee

\subsection{Rabi oscillations}

Under the action of an external alternating field, the mode fractional populations
oscillate with time as in the case of Rabi oscillations \cite{Rabi_91}. Thus for two 
modes described by equations (\ref{32}) the mode populations satisfy the equations
\be
\label{91}
\frac{dn_1}{dt} ~ = ~ 
{\rm Im} \; \left( \bt_{12}\; e^{it\Dlt\om} \; c_1^* c_2 \right) \; ,
\qquad
\frac{dn_2}{dt} ~ = ~ 
{\rm Im} \; \left( \bt_{12}^*\; e^{-it\Dlt\om} \; c_2^* c_1 \right) \; .
\ee
When $|\beta_{12}/ \alpha| \ll 1$, then the amplitudes $c_n$ can be treated as fast 
functions, while $n_1$ and $n_2$, as slow. Then, employing the averaging method, under 
the initial conditions
\be
\label{92}
 c_1(0) ~ = ~ 1 \; , \qquad c_2(0) ~ = ~ 0 \;  ,
\ee
we find \cite{Yukalov_22,Yukalov_42} the guiding centers for the populations as
\be
\label{93}
n_1 ~ = ~ 1 \; - \; 
\frac{|\;\bt_{12}\;|^2}{\Om^2} \; \sin^2\left( \frac{\Om t}{2}\right) \; , 
\qquad
n_2 ~ = ~ 
\frac{|\;\bt_{12}\;|^2}{\Om^2} \; \sin^2\left( \frac{\Om t}{2}\right) \;   ,
\ee
with the effective Rabi frequency
\be
\label{94}
 \Om^2 = [\; \Dlt\om + \al \; (n_2 - n_1 ) \; ]^2 +  |\;\bt_{12}\;|^2 \; ,
\ee
where, for simplicity, it is assumed that $\alpha_{12} = \alpha_{21} \equiv \alpha$.    
The Rabi oscillations can be used for studying the parameters of a two-level system.

\subsection{Higher-order resonances}

In addition to the resonance condition $\omega = \omega_{21}$, the generation of 
coherent modes can also be realized at higher-order resonances, under the effects
of harmonic generation, employing a single external pumping field, with
$$
n\om ~= ~\om_{21} \qquad ( n = 1,2,\ldots ) \; ,
$$
and at parametric conversion, when several fields $\omega_j$ are used, such that
$$
 \sum_j (\pm \om_j) ~= ~\om_{21} \;  .
$$ 
This can be shown \cite{Yukalov_68,Yukalov_70} by solving Eqs. (\ref{32}) using the 
scale-separation approach \cite{Yukalov_1998,Yukalov_2000}.

\section{Ramsey fringes}

Ramsey \cite{Ramsey_92} improved upon Rabi's method by splitting one interaction zone 
into two short interaction zones of time $\tau$, by applying two consecutive pulses 
separated by a long time interval $T \gg \tau$. The idea is to study the population 
$n_2$ of the excited atoms as a function of detuning or of separation time. In the 
case of coherent modes, the population $n_2$ describes the fraction of atoms in an 
excited collective coherent mode.

For a two-mode situation, equations (\ref{32}) cannot be solved exactly because of 
the nonlinearity caused by atomic interactions. However, when 
$|\beta_{12}/\alpha| \ll 1$, it is possible to use averaging techniques, since then 
$c_n$ is classified as fast, while $n_1$ and $n_2$, as slow. Then equations (\ref{32})
can be solved keeping the populations as quasi-integrals of motion. These equations
can be rewritten \cite{Yukalov_22} as
$$
\frac{d^2 c_1}{dt^2} \; - \; i (\Dlt\om - \al ) \; \frac{dc_1}{dt} +
\left[ \; \frac{|\;\bt_{12}\;|^2}{4} + 
\al_{12} n_2 \; ( \Dlt\om - \al_{21} n_1 ) \; \right]\; c_1 ~ = ~ 0 \; ,
$$
\be
\label{95}
\frac{d^2 c_2}{dt^2} + i (\Dlt\om + \al ) \; \frac{dc_2}{dt} +
\left[ \; \frac{|\;\bt_{12}\;|^2}{4} -
\al_{21} n_1 \; ( \Dlt\om + \al_{12} n_2 ) \; \right]\; c_2  ~ = ~ 0 \; .
\ee
Introducing the notation for an effective detuning
\be
\label{96}
\Dlt ~ \equiv ~ \Dlt\om + \al_{12} n_2 - \al_{21} n_1   
\ee
and for an average interaction parameter
\be
\label{97}
\al ~ \equiv ~  \al_{12} n_2 + \al_{21} n_1 \;  ,
\ee
we find the solutions of these equations
$$
c_1(t) ~ = ~ \left\{ \; 
c_1(0)\; \left[ \; \cos\left( \frac{\Om t}{2}\right) -
i\; \frac{\Dlt}{\Om}\; \sin\left( \frac{\Om t}{2}\right)\; \right] \; - \right.
$$
$$
\left.
- \;
i \; \frac{\bt_{12}}{\Om} \; c_2(0) \sin\left( \frac{\Om t}{2}\right) \; 
\right\} 
\; \exp \left\{ \; \frac{i}{2} \; (\Dlt\om - \al) \; t \; \right\} \; ,
$$
$$
c_2(t) ~ = ~ \left\{ \; 
c_2(0)\; \left[ \; \cos\left( \frac{\Om t}{2}\right) -
i\; \frac{\Dlt}{\Om}\; \sin\left( \frac{\Om t}{2}\right)\; \right] \; - \right.
$$
\be
\label{98}
- \; \left.
i \; \frac{\bt_{12}^*}{\Om} \; c_1(0) \sin\left( \frac{\Om t}{2}\right) \; 
\right\}
\; \exp \left\{ \; -\; \frac{i}{2} \; (\Dlt\om + \al) \; t \; \right\} \; .
\ee

If the first pulse acts during the time interval $0 \leq t \leq \tau$ then at the 
end of the pulse, starting from the initial conditions (\ref{92}), the solutions 
are
$$
c_1(\tau) ~ = ~ \left[\; \cos\left( \frac{\Om\tau}{2}\right) - 
i \; \frac{\Dlt}{\Om} \; \sin\left( \frac{\Om\tau}{2}\right) \; \right] \;
\exp\left\{ \; \frac{i}{2} \; ( \Dlt\om - \al) \tau \; \right\} \; ,
$$
\be
\label{99}    
c_2(\tau) ~ = ~  - i \; \frac{\bt_{12}^*}{\Om} \; 
\sin\left( \frac{\Om\tau}{2}\right) \; 
\exp\left\{ \; -\; \frac{i}{2} \; ( \Dlt\om + \al) \tau \; \right\} \; .
\ee

Then, during the time interval $\tau\leq t\leq t+T$, the external pulses are 
absent, which implies zero $\beta_{12}$. In that interval, the solutions have the 
form (\ref{98}), where the initial conditions $c_n(0)$ are replaced by $c_n(\tau)$ 
from (\ref{99}). Then, at time $\tau + T$, we have
$$
c_1(\tau+T) ~ = ~ c_1(\tau) \; 
\left[\; \cos\left( \frac{\Om(\tau+T)}{2}\right) - 
i\; \frac{\Dlt}{\Om} \; \sin\left( \frac{\Om(\tau+T)}{2}\right) \; \right] \; \times
$$
$$
\times \;
\exp\left\{ \; \frac{i}{2} \; ( \Dlt\om - \al) (\tau +T) \; \right\} \; ,
$$
$$
c_2(\tau+T) ~ = ~ c_2(\tau) \; 
\left[ \;  \cos\left( \frac{\Om(\tau+T)}{2}\right) -
i \; \frac{\Dlt}{\Om} \; \sin\left( \frac{\Om(\tau+T)}{2}\right) \; \right] \;
\times
$$
\be
\label{100}
\times \;
\exp\left\{ \; -\; \frac{i}{2} \; ( \Dlt\om + \al) (\tau +T) \; \right\} \; .
\ee
 
Following this way, at the end of the second pulse we find the amplitude of the excited  
mode
$$
c_2(2\tau+T) ~ = ~ - i \; \frac{\bt_{12}^*}{\Om} \; 
\sin\left( \frac{\Om\tau}{2} \right) \; \times
$$
$$
\times \;
\left\{ \; 
\left[ \;  \cos\left( \frac{\Om\tau}{2}\right) -
i\; \frac{\Dlt}{\Om} \; \sin\left( \frac{\Om\tau}{2}\right) \; \right] \;
\exp\left[ \; -i(\Dlt\om + \al_{12} n_2 ) \; T \; \right] \right. \; + 
$$
\be
\label{101}
+\; \left.
\left[ \;  \cos\left( \frac{\Om\tau}{2}\right) +
i \; \frac{\Dlt}{\Om} \; \sin\left( \frac{\Om\tau}{2}\right) \; \right] \;
\exp( -i \al_{21} n_1 \tau ) \; \right\} \;
\exp\{- i ( \Dlt\om +\al ) \tau \} \; .
\ee   
Therefore, the related population difference $n_2 = |c_2|^2$ reads as
$$
n_2(2\tau+T) ~ = ~ 4\; \frac{|\;\bt_{12}\;|^2}{\Om^2} \;
\sin^2\left( \frac{\Om\tau}{2}\right) \; \times
$$
\be
\label{102}
 \times \;
\left[ \; 
\cos\left( \frac{\Om\tau}{2}\right)\; \cos\left( \frac{\Dlt T}{2}\right) +
\frac{\Dlt}{\Om} \; 
\sin\left( \frac{\Om\tau}{2}\right)\; \sin\left( \frac{\Dlt T}{2}\right)\;
\right]^2 .
\ee

Usually, one works with the so-called $\pi/2$ pulses, when $\Om\tau=\pi/2$. 
Then
\be
\label{103}
 n_2(2\tau+T) ~ = ~ \frac{|\;\bt_{12}\;|^2}{\Om^2} \;
\left[ \;
 \cos\left( \frac{\Om\tau}{2}\right) + 
\frac{\Dlt}{\Om} \; \sin\left( \frac{\Dlt T}{2}\right)\;
\right]^2 \; .
\ee

The excited mode population is studied as a function of the detuning $\Dlt\om$.
The corresponding oscillating function $n_2$, demonstrating Ramsey fringes, 
is shown in Refs. \cite{Ramos_93,Ramos_94}. The Ramsey fringes, caused by coherent 
modes, are similar to Ramsey fringes of binary Bose condensate mixtures of two 
different components \cite{Eschmann_95} or of mixtures with two different internal 
states \cite{Doring_96}, or of mixtures consisting of atoms and molecules 
\cite{Kokkelmans_97}.

\section{Interaction modulation}

The atomic scattering length can be varied in the presence of an external magnetic 
field $B$ due to Feshbach resonance \cite{Timmermans_98}. The magnetic field can 
be varied in time, $B = B(t)$. Then the related scattering length 
\be
\label{104}
 a_s(B) ~ = ~ a_s\; \left( 1 \; - \; \frac{\Dlt B}{B-B_{res} } \right) \; ,
\ee
where $a_s$ is the scattering length far outside of the resonance field $B_{res}$
and $\Delta B$ is the resonance width, also becomes time dependent, as well as the 
interaction strength
\be
\label{105}
\Phi(t) ~ = ~ 4\pi \; \frac{a_s(B)}{m} \; , \qquad
B ~ = ~ B(t) \;  .
\ee
If the magnetic field $B(t)$ is modulated around a field $B_0$, so that
\be
\label{106}
B(t) ~ = ~ B_0 + b(t) \;   ,
\ee
with a small modulation amplitude $|b(t)/B_0| \ll 1$ oscillating as
\be
\label{107}
 b(t) ~ = ~ b_1\cos(\om t) + b_2 \sin(\om t) \;  ,
\ee
then the effective interaction strength acquires the form
\be
\label{108}
 \Phi(t) ~ = ~ \Phi_0 + \Phi_1 \cos(\om t) + \Phi_2 \sin(\om t) \;  ,
\ee
where
$$
\Phi_0 ~ = ~ \frac{4\pi}{m} \; a_s 
\left( 1 \; - \; \frac{\Dlt B}{B_0-B_{res}} \right) \; ,
$$
$$
\Phi_1 ~ = ~ \frac{4\pi a_s b_1 \Dlt B}{m(B_0 - B_{res})^2} \; ,
\qquad
\Phi_2 ~ = ~  \frac{4\pi a_s b_2 \Dlt B}{m(B_0 - B_{res})^2} \;  .
$$

Resonance modulation of the effective interaction can generate coherent modes, 
similarly to the trap modulation, as has been mentioned in \cite{Yukalov_99} and
studied in \cite{Yukalov_15,Ramos_100,Yukalov_101}.

Thus we are looking for the solution of the nonlinear Schr\"{o}dinger equation 
\be
\label{109}
 i\; \frac{\prt}{\prt t} \; \vp(\br,t) ~ = ~ \hat H(t) \; \vp(\br,t) \; , 
\ee    
with the effective time-dependent Hamiltonian
\be
\label{110}
\hat H(t)  ~ = ~ - \; \frac{\nabla^2}{2m} + U(\br) +
 N \; \Phi(t) \; |\; \vp(\br,t) \; |^2  \; .
\ee
Substituting here the mode expansion
\be
\label{111}
 \vp(\br,t)  ~ = ~ \sum_n c_n(t) \; e^{-iE_n t} \; \vp_n(\br) \; ,
\ee
keeping in mind the two-mode case, following the same procedure as above, and 
using the notation
$$
\gm_1  ~ \equiv ~ (\Phi_1 - i \Phi_2) \; N 
\int \vp_1^*(\br) |\; \vp_1(\br,t) \; |^2 \vp_2(\br) \; d\br \; ,
$$
\be
\label{112}
\gm_2  ~ \equiv ~ (\Phi_1 - i \Phi_2) \; N 
\int \vp_1^*(\br) |\; \vp_2(\br,t) \; |^2 \vp_2(\br) \; d\br \;   ,
\ee
we obtain the equations for the mode amplitudes
$$
i \;\frac{dc_1}{dt} ~ = ~ \al_{12} |\; c_2 \; |^2 c_1 + 
\left( 
\gm_1 \; |\; c_1 \; |^2 + \frac{1}{2} \; \gm_2 \; |\; c_2 \; |^2 \right) \; 
c_2 e^{i\Dlt\om t} + 
\frac{1}{2} \; \gm_1^* \; c_2^* \; c_1^2 \; e^{-i\Dlt\om t} \; ,
$$
\be
\label{113}
i \;\frac{dc_2}{dt} ~ = ~ \al_{21} |\; c_1 \; |^2 c_2 + 
\left( 
\gm_2^* \; |\; c_2 \; |^2 + \frac{1}{2} \; \gm_1^* \; |\; c_1 \; |^2 
\right) \; 
c_1 e^{-i\Dlt\om t} + 
\frac{1}{2} \; \gm_2 \; c_1^* \; c_2^2 \; e^{i\Dlt\om t} \; .
\ee

Generally, it is possible to combine the mode generation by means of trap modulation 
as well as by scattering length modulation. This can be used for generating multiple 
coherent modes or for enhancing the resonant generation of a special mode. 
Then, employing the same procedures as above, we derive \cite{Yukalov_101} the 
equations
$$
i \;\frac{dc_1}{dt} ~ = ~ \al_{12} |\; c_2 \; |^2 c_1 + 
\frac{1}{2} \; \left( 2\gm_1 \; |\; c_1 \; |^2 + \gm_2 \; |\; c_2 \; |^2
+\bt_{12} \right) \; c_2 \; e^{i\Dlt\om t} +
\frac{1}{2}\; \gm_1^* \; c_2^* \; c_1^2 \; e^{-i\Dlt\om t} \; ,
$$
\be
\label{114}
i \;\frac{dc_2}{dt} ~ = ~ \al_{21} |\; c_1 \; |^2 c_2 + 
\frac{1}{2} \; \left( 2\gm_2^* \; |\; c_2 \; |^2 + \gm_1^* \; |\; c_1 \; |^2
+\bt_{12}^* \right) \; c_1 \; e^{ - i\Dlt\om t} +
\frac{1}{2}\; \gm_2 \; c_1^* \; c_2^2 \; e^{i\Dlt\om t} \; .
\ee
    
Equations (\ref{113}) and (\ref{114}) have the structure similar to Eqs. (\ref{32}), 
hence they can produce the same effects involving coherent modes 
\cite{Yukalov_15,Ramos_100,Yukalov_101}. The generation of coherent modes can be 
effectively done by both methods, by means of external potential modulation or by
modulating atomic interactions 
\cite{Yukalov_15,Ramos_100,Yukalov_101,Nguyen_102,Shukono_103}.

\section{Strong interactions and noise}

In the above sections, generation of coherent modes is treated in the regime of 
asymptotically weak interactions and in the absence of noise corresponding to 
temperature. The condensate wave equation (\ref{15}) has been obtained assuming that
all atoms are condensed. This has allowed us, for deriving the condensate wave-function 
equation, to resort to the coherent approximation, resulting in the nonlinear 
Schr\"{o}dinger equation (\ref{15}). Of course, at finite, and moreover under strong 
interactions, the condensate can be essentially depleted. In addition, there can exist
intrinsic noise imitating temperature effects.

Thus, three questions arise. First, what is the wave-function equation for finite and 
strong atomic interactions? Second, is it feasible to generate coherent modes with that
condensate equation? And, third, what is the influence of intrinsic noise on the mode 
dynamics? This section provides answers to these questions. Below, we follow the 
self-consistent theory of Bose-condensed systems \cite{Yukalov_15,Yukalov_16,Yukalov_23,
Yukalov_104,Yukalov_105,Yukalov_106,Yukalov_107}.

When in a Bose system, described by the Hamiltonian (\ref{3}), there occurs 
Bose-Einstein condensation, the global gauge symmetry becomes broken. The symmetry 
breaking is the necessary and sufficient condition for Bose condensation 
\cite{Yukalov_12,Yukalov_15}. The explicit and convenient way of symmetry breaking 
is by means of the Bogolubov shift
\be
\label{115}
 \psi(\br,t) ~ = ~ \eta(\br,t) + \psi_1(\br,t) \;  ,
\ee
where $\eta({\bf r},t)$ is the condensate wave function and $\psi_1({\bf r},t)$ is the
field operator of uncondensed atoms. The field variables of condensed and uncondensed 
particles are mutually orthogonal,
\be
\label{116}
 \int \eta^*(\br,t) \; \psi_1(\br,t) \; d\br ~ = ~ 0 \;  .
\ee

The condensate wave function is normalized to the number of condensed atoms
\be
\label{117}
N_0 ~ = ~ \int |\; \eta(\br,t) \; |^2 \; d\br \;   ,
\ee
while the number of uncondensed atoms is 
\be
\label{118}
N_1 ~ = ~ \int \lgl \; \psi_1^\dgr(\br,t) \; \psi_1(\br,t) \; \rgl \; d\br \;   .
\ee
The total number of atoms
\be
\label{119}
 N ~ = ~ N_0 + N_1
\ee   
is fixed. 

In order that the quantum numbers, such as spin or momentum, be conserved, requires
to impose the conservation condition
\be
\label{120}
\lgl \;  \psi_1(\br,t) \; \rgl  ~ = ~ 0 \; .
\ee
The general form of the conservation condition is written as
\be
\label{121}
 \lgl \;  \hat \Lbd \; \rgl  ~ = ~ 0 \;  ,
\ee
with the condition operator
\be
\label{122}
 \hat\Lbd ~ = ~ \int \left[\; \lbd(\br,t) \; \psi_1^\dgr(\br,t) + 
\lbd^*(\br,t) \; \psi_1(\br,t) \; \right] \; d\br \; ,
\ee
in which $\lambda({\bf r},t)$ is a Lagrange multiplier. 

Taking into account the normalization conditions (\ref{117}) and (\ref{118}), as well 
as the conservation condition (\ref{121}), leads to the grand Hamiltonian
\be
\label{123}
H ~ = ~ \hat H - \mu_0 N_0 - \mu_1 \hat N_1 - \hat\Lbd \;   ,
\ee
where
\be
\label{124}
  \hat N_1 ~ = ~ \int  \psi_1^\dgr(\br,t) \;  \psi_1(\br,t) \; d\br \; .
\ee

The evolution equations for condensed and uncondensed atoms are
\be
\label{125}
i\; \frac{\prt}{\prt t} \; \eta(\br,t) ~ = ~ 
\left\lgl \; \frac{\dlt H}{\dlt\eta^*(\br,t)} \; \right\rgl
\ee
and, respectively,
\be
\label{126}
i\; \frac{\prt}{\prt t} \; \psi_1(\br,t) ~ = ~ 
\frac{\dlt H}{\dlt\psi_1^\dgr(\br,t)} \;  .
\ee
The conservation condition (\ref{121}) requires that the grand Hamiltonian must have 
no linear in $\psi_1$ terms, which is achieved by setting the Lagrange multiplier
\be
\label{127}
\lbd(\br,t) ~ = ~ \left[\; - \; \frac{\nabla^2}{2m} + U(\br,t) +
\frac{1}{2} \int \Phi(\br-\br') \; |\; \eta(\br',t) \; |^2 \; d\br' \; \right] \;
\eta(\br,t) \;  .
\ee

To proceed further, let us introduce several notations, such as the normal density 
matrix  
\be
\label{128}
 \rho_1(\br,\br',t) ~ \equiv ~ 
\lgl \;  \psi_1^\dgr(\br',t) \;  \psi_1(\br,t) \; \rgl \; ,
\ee
the so-called anomalous average
\be
\label{129}
\sgm_1(\br,\br',t) ~ \equiv ~ 
\lgl \;  \psi_1(\br',t) \;  \psi_1(\br,t) \; \rgl \;  ,
\ee
the density of condensed atoms
\be
\label{130}
 \rho_0(\br,t) ~ \equiv ~ |\; \eta(\br,t) \; |^2 \;  ,
\ee
the density of uncondensed atoms
\be
\label{131}
 \rho_1(\br,t) ~ \equiv ~ 
\lgl \;  \psi_1^\dgr(\br,t) \;  \psi_1(\br,t) \; \rgl ~ = ~
\rho_1(\br,\br,t) \;   ,
\ee
the total density of atoms
\be
\label{132}
 \rho(\br,t) ~ = ~ \rho_0(\br,t) + \rho_1(\br,t) \;  ,
\ee
the diagonal anomalous average
\be
\label{133}
\sgm_1(\br,t) ~ \equiv ~ 
\lgl \;  \psi_1(\br,t) \;  \psi_1(\br,t) \; \rgl ~ = ~
\sgm_1(\br,\br,t) \;   ,
\ee
and the triple anomalous average
\be
\label{134}
\xi(\br,\br',t) ~ \equiv ~ 
\lgl \;   \psi_1^\dgr(\br',t) \; \psi_1(\br',t) \;  \psi_1(\br,t) \; \rgl \;   .
\ee

With these notations, the equation for the condensate wave function reads as
$$
i\; \frac{\prt}{\prt t} \; \eta(\br,t) ~ = ~ 
\left[ -\; \frac{\nabla^2}{2m} + U(\br,t) - \mu_0 \; \right] \; \eta(\br,t) \; + 
$$
\be
\label{135}
+ \;
\int \Phi(\br-\br') \; \left[ \;
\rho(\br',t) \; \eta(\br,t) + \rho_1(\br,\br',t) \; \eta(\br',t) +
\sgm(\br,\br',t) \; \eta^*(\br',t) + \xi(\br,\br',t) \; \right] \; d\br' \; .
\ee

Note that this is an exact equation valid for any integrable interaction potential.
For the contact potential (\ref{4}), the equation takes the form
$$
i\; \frac{\prt}{\prt t} \; \eta(\br,t) ~ = ~ 
\left[ -\; \frac{\nabla^2}{2m} + U(\br,t) - \mu_0 \; \right] \; \eta(\br,t) \; +
$$
\be
\label{136}
+\;
\Phi_0 \{ \; [\; \rho_0(\br,t) + 2\rho_1(\br,t) \; ] \; \eta(\br,t) +
\sgm_1(\br,t) \; \eta^*(\br,t) + \xi(\br,t) \; \} \;  ,
\ee
in which
\be
\label{137}
 \xi(\br,t) ~ \equiv ~ \xi(\br,\br,t) \;  .
\ee
  
The solution to this equation can be sought in the form
\be
\label{138}
 \eta(\br,t) ~ = ~ \sqrt{N_0(t) } \; \sum_n c_n(t) \; \vp_n(\br) \;
\exp\{ - i [\; E_n + \ep_n(t) - \mu_0 \; ] \; t \} \;  .
\ee
If the number of condensed atoms changes slowly in time, so that
\be
\label{139}
\left| \; \frac{1}{E_n} \; \frac{dN_0(t)}{dt} \; \right| ~ \ll ~ 1
\ee
and the phase $\varepsilon_n(t)$ is also a slow function of time,
\be
\label{140}
 \left| \; \frac{1}{E_n} \; \frac{d\ep_n(t)}{dt} \; \right| ~ \ll ~ 1 \; ,
\ee
then, in the case of two coherent modes, we obtain \cite{Yukalov_15,Yukalov_105} 
the same equations (\ref{114}). However, the phase $\varepsilon_n(t)$ is a slow 
function when $\rho_1$, $\sigma_1$, and $\xi$ are small as compared to $\rho_0$.

In this way, when interactions between atoms are strong and temperature is high, the
phase of the condensate wave function can very rather fast. In order to understand 
how fast phase fluctuations influence the dynamics of the condensate wave function,
it is possible to consider the equations
$$
\frac{ds}{dt} ~ = ~ - \bt \; \sqrt{1 - s^2} \; \sin x \; ,
$$
\be
\label{141}
 \frac{dx}{dt} ~ = ~ \al \; s + \frac{\bt s}{\sqrt{1-s^2} } \; \cos x +
\dlt + \sgm \; dW_t \; ,
\ee
in which phase fluctuations are modeled by random noise, where $W_t$ is the standard 
Wiener process and $\sigma$ is the standard noise deviation. These equations are to be
compared with equations (\ref{51}) containing no random phase fluctuations. If the 
strength of the noise is not too large, its influence leads to the distortion of Rabi 
oscillations, but does not change the overall qualitative picture 
\cite{Yukalov_108,Yukalov_109}, as is shown in Fig. 2 for the mode-locked (subcritical) 
regime) and Fig. 3 for the mode-unlocked (supercritical regime).     

\begin{figure}[ht]
\centerline{
\hbox{ \includegraphics[width=7.5cm]{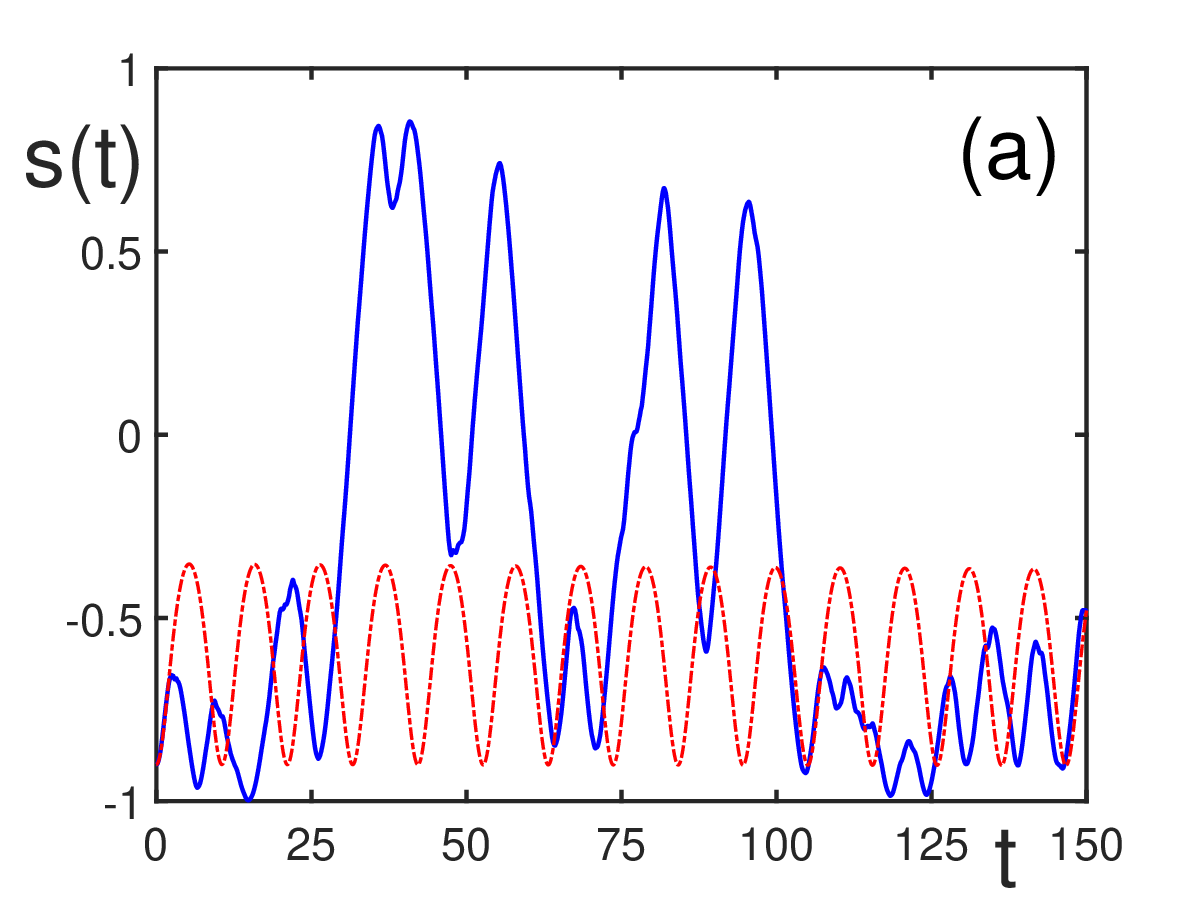} \hspace{0.5cm}
\includegraphics[width=7.5cm]{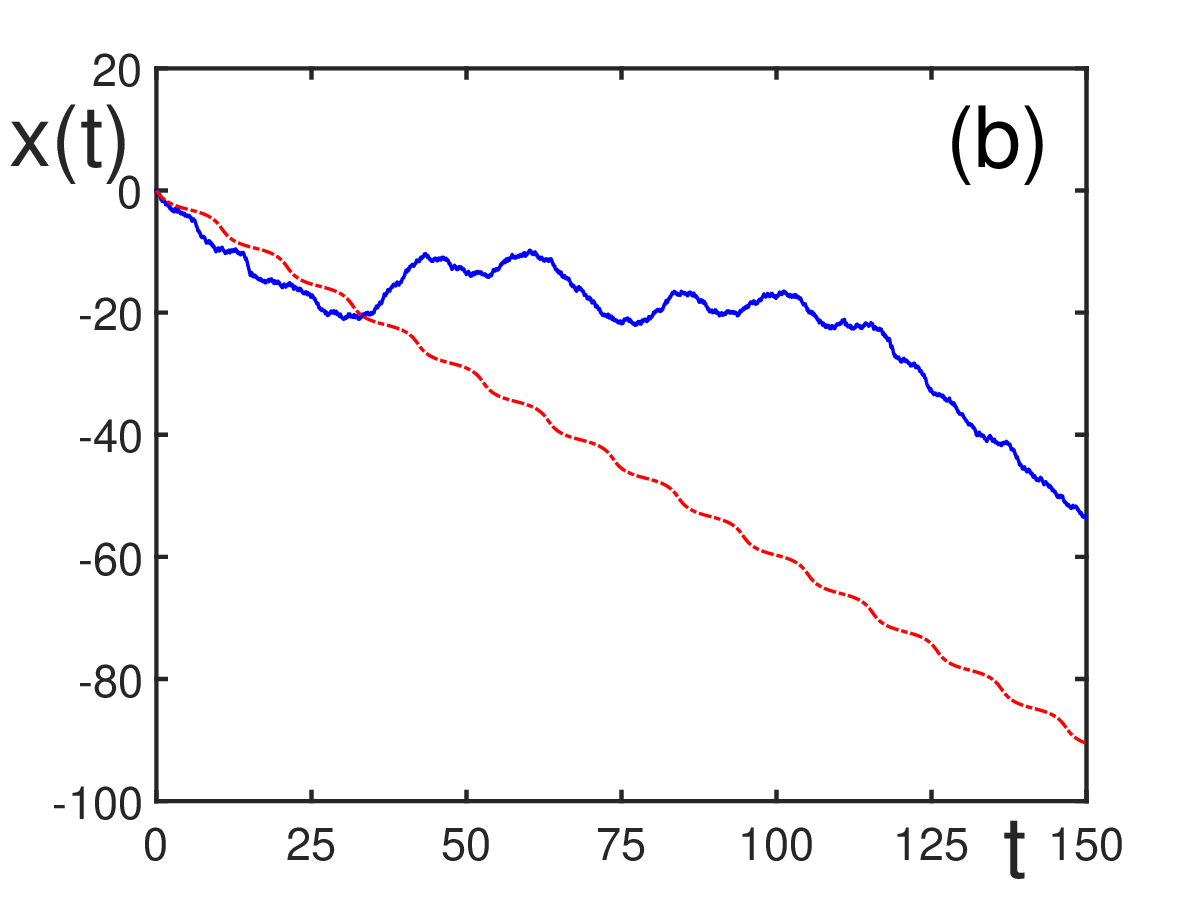}  } }
\caption{\small
Subcritical regime with $b=0.25<b_c=0.282$. 
(a) Population imbalance s(t) as functions of dimensionless time, 
for $\sgm=0.5$ (solid line) and $\sgm=0$ (dashed-dotted line).
(b) Phase difference x(t) as functions of dimensionless time, 
for $\sgm=0.5$ (solid line) and $\sgm=0$ (dashed-dotted line).
}
\label{fig:Fig.2}
\end{figure}

\begin{figure}[ht]
\centerline{
\hbox{ \includegraphics[width=7.5cm]{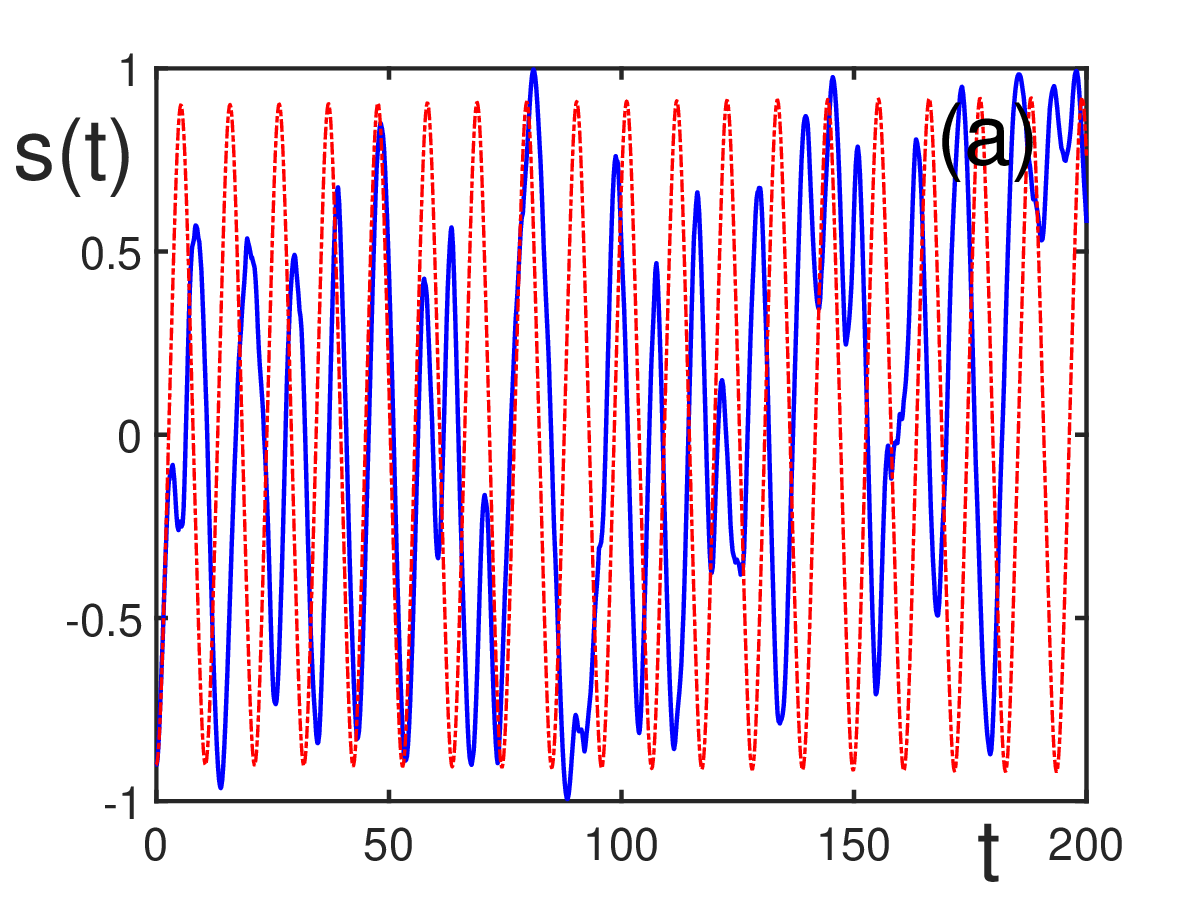} \hspace{0.5cm}
\includegraphics[width=7.5cm]{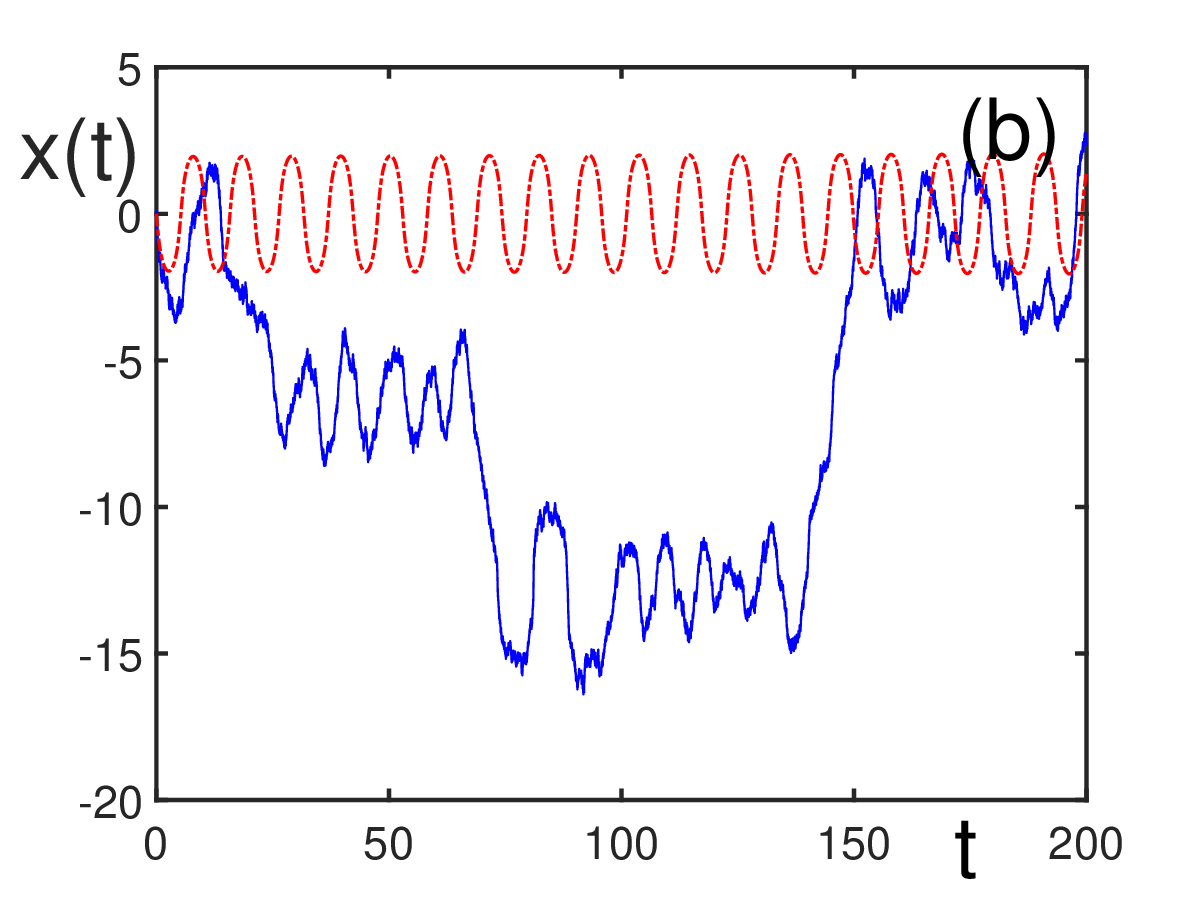}  } }
\caption{\small
Supercritical regime with $b=0.5>b_c=0.282$. 
(a) Population imbalance s(t) as functions of dimensionless time, 
for $\sgm=0.5$ (solid line) and $\sgm=0$ (dashed-dotted line).
(b) Phase difference x(t) as functions of dimensionless time, 
for $\sgm=0.5$ (solid line) and $\sgm=0$ (dashed-dotted line).
}
\label{fig:Fig.3}
\end{figure}

\section{Critical phenomena}

The qualitative change of dynamics between the mode-locked regime and mode-unlocked 
regime reminds a phase transition in a statistical system. This similarity can be 
emphasized by considering an effective stationary system obtained by averaging the 
original system over time \cite{Yukalov_41,Yukalov_64,Yukalov_65,Ramos_2007}. 

For simplicity, we set $\alpha_{12} = \alpha_{21} = \alpha$. In the case of two 
coherent modes, the evolution equations (\ref{32}) can be derived from the 
Hamiltonian equations
\be
\label{142}
i\; \frac{dc_1}{dt} ~ = ~ \frac{\prt H}{\prt c_1^*} \; ,
\qquad
i\; \frac{dc_2}{dt} ~ = ~ \frac{\prt H}{\prt c_2^*} \;   ,
\ee
with the effective Hamiltonian
\be
\label{143}
   H ~ = ~ \al \; |\; c_1 \; c_2 \; |^2 + 
\frac{1}{2} \; \left( \bt \; c_1^* \; c_2 \; e^{i\Dlt\om t} + 
\bt^* \; c_2^* \; c_1 \; e^{-i\Dlt\om t} \right) \; ,
\ee
in which $\beta \equiv \beta_{12}$. Resorting to the averaging techniques yields 
solution (\ref{98}). 

Averaging Hamiltonian (\ref{143}) over time, with using the normalization condition
\be
\label{144}
n_1 + n_2 ~ = ~ 1 \; ,
\ee
gives the effective energy
\be
\label{145}
E ~ = ~ \frac{\al|\;\bt\;|^2}{2\overline\Om^2} \; \left( 
 \frac{|\;\bt\;|^2}{2\overline\Om^2} + \ep \right) \;  ,
\ee
with $\varepsilon \equiv \delta/ \alpha$, the average Rabi frequency given by the 
expression
\be
\label{146}
\overline\Om^2 ~ = ~ \al^2 \; \left( \ep + \overline n_2 - \overline n_1 \right)^2 +
 |\;\bt\;|^2 \;  ,   
\ee
and the average fractions
\be
\label{147}
\overline n_1 ~ = ~ 1 - \overline n_2 \; , \qquad
\overline n_2 ~ = ~  \frac{|\;\bt\;|^2}{2\overline\Om^2} \;  .
\ee

The order parameter can be defined as the average population difference
\be
\label{148}
 \eta ~ \equiv ~  \overline n_1 - \overline n_2 ~ = ~ 1 \; - \; 
 \frac{|\;\bt\;|^2}{\overline\Om^2} \;  .
\ee
Also, let us introduce the system capacity to store the energy pumped into the system 
\be
\label{149}
 C_\bt ~ \equiv ~ \frac{\prt E}{\prt|\; \bt\; |}  
\ee
and the susceptibility with respect to detuning
\be
\label{150}
 \chi_\ep ~ \equiv ~ \frac{\prt \eta}{\prt|\; \ep\; |} \;  .
\ee
 
The abrupt change of dynamics happens on the critical line $b_c + \varepsilon = 0.5$,
where $b \equiv \beta/ \alpha$. In the vicinity of the dynamic phase transition, the 
system characteristics strongly depend on the variable
\be
\label{151}
 \tau ~ \equiv ~ \frac{|\; b - b_c\;|}{b_c} \;  .
\ee
When $\tau \ra 0$, we observe the critical behavior of the order parameter
\be
\label{152}
 \eta - \eta_c ~ \simeq ~ \frac{1}{\sqrt{2}} \; ( 1 - 2\ep ) \; \tau^{1/2} 
\qquad
( \eta_c ~ = ~ 0.5 + \ep ) \;  ,
\ee
effective system capacity and susceptibility:
\be
\label{153}
 C_\bt ~ \simeq ~ \frac{1}{4\sqrt{2}} \; \tau^{-1/2} \; , \qquad
 \chi_\ep ~ \simeq ~ \frac{1}{\sqrt{2}} \; \tau^{-1/2} \; .
\ee
The latter diverge as at phase transitions of second order.

The critical exponents of the order parameter, capacity, and susceptibility satisfy 
the scaling relation
\be
\label{154}
  \frac{1}{2} + 1 + \frac{1}{2} ~ = ~ 2 
\ee
typical of equilibrium second-order phase transitions.

\section{Atomic squeezing}

Bose-Einstein condensation is accompanied by global gauge symmetry breaking, when the
arising condensate is characterized by a wave function. Rigorous description of the 
Bose-condensate assumes the thermodynamic limit. In a finite volume, strictly speaking,
there is no Bose-condensate, but there exists a quasi-condensate that is described
by field operators. Hence, a finite cloud of trapped atoms should be characterized 
by operators instead of non-operator functions. In order to consider a finite 
quasi-condensate, one has to replace the amplitudes $c_n$ by Bose operators 
${\hat c}_n$. 

When one is interested in the case of two modes, one can make the canonical 
transformation to quasi-spin operators
$$
\hat c_1^\dgr \; \hat c_1 ~ = ~ \frac{1}{2} \; - \; S_z \; , 
\qquad
\hat c_2^\dgr \; \hat c_2 ~ = ~ \frac{1}{2} \; + \; S_z \; ,
$$
\be
\label{155}
\hat c_1^\dgr \; \hat c_2 ~ = ~ S_x - i S_y \; ,
\qquad
\hat c_2^\dgr \; \hat c_1 ~ = ~ S_x + i S_y \; ,
\ee
where
$$
S_\al ~ \equiv \sum_{j=1}^N S_j^\al \qquad ( \al = x,y,z ) \;   .
$$
The evolution equations can be derived \cite{Yukalov_42} using the standard Heisenberg 
equations of motion with the Hamiltonian
\be
\label{156}
H ~ = ~ - \; \frac{\al}{N} \; S_z^2 + \frac{\bt}{2} \; ( S^- + S^+ ) -
S_z \; \dlt \;   ,   
\ee
in which $S_j^\alpha$ are half-spin operators and the ladder operators are
$$
 S^\pm ~ \equiv ~ S_x \pm i S_y \;  .
$$
Averaging these equations, with employing the mean-field approximation, gives equations 
equivalent to Eqs. (\ref{51}).   

Then it is possible to study the problem of squeezing in trapped atomic systems, 
similarly to the squeezing of light in optical systems \cite{Mandel_50}. In general, 
atomic squeezing can be achieved for both bosons as well as for fermions. Squeezing 
for Bose-Einstein condensates was considered for two-component mixtures \cite{Poulsen_111},
for atoms with two internal states \cite{Sorensen_112}, and for atoms in linked mesoscopic 
traps formed by an optical lattice \cite{Orzel_113}, which is equivalent to a 
multicomponent mixture. Atomic squeezing can be used for atomic spectroscopy, creation of
atomic clocks \cite{Wineland_114}, and for atom interferometers \cite{Boyer_115}. 

For any two operators, $\hat{A}$ and $\hat{B}$, there exists the Heisenberg uncertainty
relation
\be
\label{157}
 {\rm var}(\hat A) \;  {\rm var}(\hat B) ~ \geq ~ \frac{1}{4} \;
\left| \; \lgl \; [\;\hat A, \; \hat B \;] \; \rgl \; \right|^2 \; ,
\ee
where the variance is
$$
 {\rm var}(\hat A) ~ \equiv ~ \lgl \; \hat A^+ \; \hat A \; \rgl - 
|\; \lgl \; \hat A \; \rgl \; |^2 \;   .
$$
One says that the operators $\hat{A}$ and $\hat{B}$ are not squeezed with respect to 
each other, when
\begin{eqnarray}
\label{158}
\begin{array}{l}
 {\rm var}(\hat A) ~ \geq ~ \frac{1}{2} \;
| \; \lgl \; [\;\hat A, \; \hat B \;] \; \rgl \; | \\
 {\rm var}(\hat B) ~ \geq ~ \frac{1}{2} \;
| \; \lgl \; [\;\hat A, \; \hat B \;] \; \rgl \; |
\end{array}
~~~ ( not \; squeezed) \;  .
\end{eqnarray}
And one says that $\hat{A}$ is squeezed with respect to $\hat{B}$, if
\be
\label{159}
{\rm var}(\hat A) ~ < ~ \frac{1}{2} \;
\left| \; \lgl \; [\;\hat A, \; \hat B \;] \; \rgl \; \right| \qquad
(squeezed) \;   .
\ee

The measure of squeezing is quantified by the squeezing factor
\be
\label{160}
 Q_{AB} ~ \equiv ~ 
\frac{2{\rm var}(\hat A)}{|\; \lgl\; [\;\hat A,\; \hat B \;] \; \rgl \;| } \;  .
\ee
For two operators, squeezing happens when the squeezing factor is smaller than unity,
\begin{eqnarray}
\label{161}
\begin{array}{ll}
Q_{AB} ~ \geq ~ 1 ~~ & ~~ ( no \; squeezing) \; , \\
Q_{AB} ~ < ~ 1    ~~ & ~~ (squeezing) \;  .
\end{array}
\end{eqnarray}
The occurrence of squeezing of $\hat{A}$ with respect to $\hat{B}$ implies that the 
observable corresponding to $\hat{A}$ can be measured more precisely than that one
related to $\hat{B}$.

In our case, the operator $S_z$ corresponds to the population difference, while 
$S^{\pm}$, to the transition dipoles, since
\be
\label{162}
\lgl \; S_z \; \rgl ~ = ~ \frac{N}{2} \; s \; , \qquad
 |\; \lgl\; S^\pm \; \rgl \;| ~ = ~ \frac{N}{2} \; 
|\; \lgl\; \hat c_1^\dgr \; \hat c_2 \; \rgl \;| \; .
\ee
For the related operators, the Heisenberg uncertainty relation is
\be
\label{163}
{\rm var}(S_z) \; {\rm var}(S^\pm ) ~ \geq ~ 
\frac{1}{4} |\; \lgl \; S^\pm \; \rgl \; |^2  \; .
\ee
Squeezing of $S_z$ with respect to $S^{\pm}$ is measured by the squeezing factor
\be
\label{164}
 Q ~ = ~ \frac{2{\rm var}(S_z)}{  |\; \lgl\; S^\pm \; \rgl \;| } \; ,
\ee
which leads to 
\be
\label{165}
Q ~ = ~ \sqrt{1 - s^2 } \;   .
\ee
In the mode-locked regime, $s^2 \leq 1$, and $Q < 1$ for all times. In the 
mode-unlocked regime $s^2$ for the most of time is smaller than one, touching one 
only at some moments of time. Hence almost always $S_z$ is squeezed with respect to
$S^{\pm}$. This means that the population difference, hence the mode populations,
can be measured with a better accuracy than the quantities associated with the 
transition dipoles, such as density current. 

Similarly, it is possible to consider the effect of squeezing for several coupled 
modes. Thus, for three modes, it is straightforward to introduce the quasi-spin 
operators
$$
J_+ ~ \equiv ~ \sqrt{2} \; ( a_2^\dgr \; a_1 + a_3^\dgr \; a_2 ) \; , 
\qquad
J_- ~ \equiv ~ \sqrt{2} \; ( a_1^\dgr \; a_2 + a_2^\dgr \; a_3) \; ,
\qquad
J_z ~ \equiv ~  a_3^\dgr \; a_3 - a_1^\dgr \; a_1 \;  ,
$$
satisfying the commutation relations
$$
 [\; J_+ , \; J_- \;] ~ = ~ 2 J_z \; , \qquad
 [\; J_z , \; J_\pm \;] ~ = ~ \pm J_\pm \;  .
$$
Then the squeezing factor is defined in the same way as for two modes.

\section{Cloud entanglement}

It is necessary to distinguish two different notions, {\it entanglement of states} 
and {\it entanglement production}. Entanglement of a state concerns the structure
of a statistical operator, while entanglement production by an operator describes
the ability of an operator, by acting on functions of a Hilbert space, to create 
entangled functions.

\subsection{Entangled states}

When one talks about entanglement, one has, first of all, to specify what are the 
objects whose entanglement is considered. In quantum physics it is customary to 
consider entanglement of particles. However in the case of Bose-Einstein condensate,
it is possible to study entanglement between large atomic clouds. Such coherent atomic
clouds can be organized in optical lattices with sufficiently deep wells at lattice 
sites. Each lattice well can house condensate clouds containing between several atoms
to $10^4$ atoms \cite{Hadzibabic_116,Cennini_117}. Shaking the optical lattice by means 
of magnetic fields or lasers, it is possible to excite coherent modes simultaneously 
in all lattice sites.

Suppose Bose-condensed atoms are loaded into a deep lattice with $N_L$ sites, enumerated
by the index $j = 1,2,\ldots, N_L$, where at each lattice site there is a cloud of $N$ 
atoms, with $N \gg 1$. Assume that there can be excited $M$ coherent modes enumerated 
by the index $m = 1, 2, \ldots, M$. A coherent mode $m$ at a lattice site $j$, which is
characterized by a coherent field $\eta_{jm}({\bf r})$, in the Fock space is represented
\cite{Yukalov_118} by a column
\be
\label{166}
| \; \eta_{jm} \; \rgl ~ = ~ \left[ \; \frac{e^{-N/2}}{\sqrt{n!} } \;
\prod_{i=1}^n \eta_{jm}(\br_i) \; \right] \;   ,
\ee
with the rows enumerated by $n = 0,1,2, \ldots$. These modes form a basis in the space
\be
\label{167}
\cH_j ~ = ~ {\rm span}_m \{ \; |\; \eta_{jm} \; \rgl \; \} \; .
\ee
The overall system of $N_L$ clouds is described by the Hilbert space
\be
\label{168}
\cH ~ = ~ \bigotimes_{j=1}^{N_L} \cH_j ~ = ~ 
{\rm span}_m \left\{  
\bigotimes_{j=1}^{N_L} |\; \eta_{jm} \; \rgl \; \right\} \;  .
\ee
The coherent wave state of the whole system in the Fock space is
\be
\label{169}
 |\; \eta (t)\; \rgl ~ = ~ \sum_{m=1}^M 
c_m(t) \bigotimes_{j=1}^{N_L} |\; \eta_{jm} \; \rgl \;  .
\ee
The related statistical operator, or statistical state, is
\be
\label{170}
\hat\rho(t) ~ = ~   |\; \eta (t)\; \rgl  \lgl \; \eta (t) \; | \; .
\ee
Notice that the wave state (\ref{169}) is a multimode multicat entangled state.  

The structure of the state (\ref{170}), 
\be
\label{171}
\hat\rho(t) ~ = ~ \sum_{m=1}^M \sum_{n=1}^M c_m(t) \; c_n^*(t) 
 \bigotimes_{j=1}^{N_L} |\; \eta_{jm} \; \rgl  \lgl \; \eta_{jn} \; | \; ,
\ee
shows that this is an entangled state, with respect to clouds located at the lattice 
wells \cite{Yukalov_119}, provided that there exist at least two nontrivial modes,
when $M \geq 2$. The state (\ref{170}) is entangled, since it is not separable 
\cite{Keyl_120}, hence cannot be represented in the form
\be
\label{172}
\hat\rho_{sep} ~ = ~ \sum_k \lbd_k \bigotimes_{j=1}^{N_L} \hat\rho_{jk}
\qquad
\left( \lbd_k \geq 0 , \; \sum_k \lbd_k = 1 \right) \; ,
\ee
where $\hat{\rho}_{jk}$ are statistical operators in the spaces (\ref{167}).

\subsection{Entanglement production}

Entanglement production by an operator characterizes the ability of this operator 
of entangling the functions of the Hilbert space it acts on. An operator is called 
{\it entangling}, if there exists at least one separable pure state such that it 
becomes entangled under the action of the operator.

Suppose an operator $\hat{A}$ acting on a Hilbert space 
$\mathcal{H} = \otimes_j \mathcal{H}_j$ is a trace-class operator with a finite nonzero 
trace,
\be
\label{173}
0 ~ < ~ |\; {\rm Tr} \hat A \; | ~ < ~ \infty \;   .
\ee
Generally, an operator, acting on a state, pertaining to $\mathcal{H}$, produces 
an entangled state, even when this operator is separable, being defined similarly 
to (\ref{172}) as
\be
\label{174}
 \hat A_{sep}  ~ = ~ 
\sum_k \lbd_k \bigotimes_j \hat A_{jk} 
\qquad 
\left( \lbd_k \geq 0 , \; \sum_k \lbd_k = 1 \right) \;  .
\ee
Even when it acts on a product state, it generates an entangled state
\be
\label{175}
\hat A_{sep} \bigotimes |\; \vp_j \; \rgl ~ = ~ 
\sum_k \lbd_k \bigotimes_j \hat A_{jk} \; | \; \vp_j \; \rgl \;   ,
\ee
if at least two $\lambda_k$ differ from zero. 

The sole operator that does not entangle product states is the operator of the factor 
form
\be
\label{176}
 \hat A^\otimes ~ = ~ \bigotimes_j \hat A_j \;   ,
\ee
which is clear from its action
\be
\label{177}  
\hat A^\otimes \bigotimes_j |\; \vp_j \; \rgl  ~ = ~ 
\bigotimes_j \hat A_j \; |\; \vp_j \; \rgl  .
\ee

The measure of entanglement production has been introduced 
\cite{Yukalov_121,Yukalov_122,Yukalov_123} by comparing the action of $\hat{A}$ with
that of its nonentangling counterpart
\be
\label{178}
 \hat A^\otimes ~ \equiv ~ 
\frac{\bigotimes_j \hat A_j}{({\rm Tr}_\cH \hat A)^{N_L-1} } \;  ,
\ee
where
\be
\label{179}
\hat A_j ~ \equiv ~ {\rm Tr}_{\cH/\cH_j} \; \hat A    
\ee
and the normalization holds:
\be
\label{180}
  {\rm Tr}_{\cH} \; \hat A^\otimes ~ = ~  {\rm Tr}_{\cH} \; \hat A \; .
\ee

The measure of entanglement production is defined \cite{Yukalov_121,Yukalov_122} as
\be
\label{181}
\ep(\hat A) ~ \equiv ~ 
\log \; \frac{||\; \hat A\; ||}{||\; \hat A^\otimes\; || } \;   ,
\ee
where any logarithm base can be accepted. This measure is semi-positive, zero for 
nonentangling operators, continuous, additive, and invariant under local unitary 
operations \cite{Yukalov_121}. 

The fact that the only nontrivial operators preserving separability are the operators 
having the form of tensor products of local operators is proved in Refs. 
\cite{Marcus_124, Westwick_125,Beasley_126,Alfsen_127,Johnston_128,Friedland_129}.
The operator preserving separability is called nonentangling \cite{Gohberg_130}.
The operator transforming at least one non-entangled state into an entangled state 
is termed entangling \cite{Fan_131}.

\subsection{Resonant entanglement production}

Existence of entanglement in a statistical operator implies the occurrence of quantum 
correlation between subsystems. Then the statistical average of the operator of an
observable cannot be factorized into partial averages. This kind of factorization, 
reminding a mean-field approximation, happens only when both, the operator of an 
observable, as well as the statistical operator are non-entangling, so that
\be
\label{182}
 {\rm Tr}_\cH \; \hat\rho \bigotimes_j \hat A_j ~ = ~ \prod_j 
{\rm Tr}_{\cH_j} \; \hat\rho_j \; \hat A_j \qquad 
( \hat\rho = \otimes_j \hat\rho_j ) \;  ,
\ee
which can be represented as
\be
\label{183}
\lgl \; \bigotimes_j A_j \; \rgl ~ = ~ \prod_j \lgl \; \hat A_j \; \rgl \; .
\ee

The statistical operator (\ref{171}) for a system of atomic clouds, with resonantly 
generated coherent modes, is entangled, so that the considered above factorization
cannot happen. The statistical operator (\ref{171}) is also entangling. Thus its 
action on the non-entangled coherent basis state of space (\ref{168}) yields an
entangled state
\be
\label{184}
\hat\rho(t) \bigotimes_{j=1}^{N_L} |\; \eta_{jm} \; \rgl ~ = ~
c_m^*(t) \sum_n c_n(t) \bigotimes_{j=1}^{N_L} |\; \eta_{jn} \; \rgl \;   .
\ee 

The entangling power of this statistical operator is quantified \cite{Yukalov_132}
by the measure of entanglement production
\be
\label{185}
 \ep(\hat\rho(t)) ~ = ~ 
\log \; \frac{||\;\hat\rho(t)\;||}{||\;\hat\rho^\otimes(t)\;||} \; ,
\ee
in which
\be
\label{186}
\hat\rho^\otimes ~ = ~ \bigotimes_{j=1}^{N_L} \hat\rho_j(t) \; , \qquad
\hat\rho_j(t) ~ \equiv ~ {\rm Tr}_{\cH/\cH_j} \; \hat\rho(t) \;  .
\ee
Using the standard operator norm gives the entanglement measure produced by the 
statistical operator (\ref{171}) as
\be
\label{187}
\ep(\hat\rho(t)) ~ = ~ 
( 1 - N_L) \; \log \; \sup_m \; |\; c_m(t) \; |^2 \;   .
\ee
 
Creation of entanglement in a lattice of atomic clouds with resonantly generated 
coherent modes can be called {\it resonant entanglement production}. The behavior 
of measure (\ref{187}) has been studied in Refs. \cite{Yukalov_132,Yukalov_133}. 
A number of other examples of entanglement production are considered in Refs. 
\cite{Yukalov_134,Yukalov_135,Yukalov_136}.

\section{Nonresonant mode generation}

As is stressed above, coherent modes can be excited by modulating either the 
trapping potential or the scattering length. There are also two ways of generation, 
resonant and nonresonant. In the case of resonant generation, the frequency of the 
alternating field has to be in resonance with the transition frequency corresponding 
to the transition between the ground-state energy level and the energy of the mode 
to be generated. Then the amplitude of modulation by an external field does not need 
to be very large. 

However the strict occurrence of the resonance is not compulsory for mode generation 
due to the effect of power broadening, when modes can be generated if the modulation 
frequency is not in resonance with the transition frequency, but the alternating 
field amplitude is sufficiently strong to generate higher-order coherent modes. 
Depending on the strength of the alternating field, those modes are excited for which 
the energy pumped by the field is sufficient for their creation.      

If the aim would be to generate quantum vortices, then it could be done by means of 
a rotating laser beam. However, if one wishes to generate different kinds of modes,  
then there is no reason of imposing anisotropy on the system, so that the alternating
potential should be more or less isotropic.   

Thus in experiments \cite{Henn_137,Seman_138} the harmonic trapping potential
\be
\label{188}
 U(\br) ~ = ~ 
\frac{m}{2} \; \om_x^2 \; ( x^2 + y^2 ) + \frac{m}{2} \; \om_z^2 \; z^2 \;  ,
\ee
with the trapping frequencies $\om_x=2\pi\times 210$ Hz and $\om_z=2\pi\times 23$ Hz, 
is subject to the alternating field with the frequency $\omega =  2 \pi \times 200$ Hz 
and the modulating amplitude $0.2 \omega_x$. The modulated potential has the form
\be
\label{189}
U(\br,t) ~ = ~ \frac{m}{2} \; \Om_x^2(t) \; ( x - x_0)^2 +
\frac{m}{2} \; \Om_y^2(t) \; ( y' - y_0')^2 +
\frac{m}{2} \; \Om_z^2(t) \; ( z' - z_0')^2 \; ,
\ee
where the oscillation centers are given by the equation
\begin{eqnarray}
\nonumber
\left[\; \begin{array}{l}
y' - y_0' \\
z' - z_0' \end{array} \right] ~ = ~
\left[\; \begin{array}{rr}
\sin\vartheta_0 ~ & ~ \cos\vartheta_0 \\
\cos\vartheta_0 ~ & ~ -\sin\vartheta_0 \end{array} \right] \;
\left[\; \begin{array}{l}
y - y_0 \\
z - z_0 \end{array} \right]
\end{eqnarray}
and the oscillation frequencies are
$$
\Om_\al(t) ~ = ~ \om_\al \; \dlt_\al \; [\;  1 - \cos(\om t)  \; ]  \; .
$$
By increasing the strength of the alternating potential and the modulation time, it 
is possible to pump into the trap large amounts of energy. The energy, per atom, 
injected into the trap during the period of time $t$ can be written as 
\be
\label{190}
E_{inj} ~ = ~ 
\frac{1}{N} \int_0^t \left| \; \frac{\prt H(t)}{\prt t} \; \right| \; dt \;  ,
\ee
where
\be
\label{191}
 H(t) ~ = ~ 
\int \eta^*(\br,t) \; \left[ \; - \; \frac{\nabla^2}{2m} + U(\br,t) \; \right] \; 
\eta(\br,t) \; d\br + 
\frac{1}{2} \; \Phi_0 \int |\; \eta(\br,t) \; |^4 \; d\br \; .
\ee
This type of modulation imposes no anisotropy on the system.

The overall procedure of generating coherent modes and different nonequilibrium states,
by employing the above alternating potential, has been studied in detail by realizing 
experimental observations, providing theoretical description, and accomplishing numerical 
calculations \cite{Henn_139,Shiozaki_140,Seman_141,Bagnato_142,Yukalov_143,Yukalov_144,
Yukalov_145,Novikov_146,Yukalov_147,Yukalov_148,Yukalov_149}. 

The same setup has been used in numerical simulations as well as in experiments 
with $^{87}$Rb \cite{Henn_137,Seman_138}. At the initial moment of time, practically 
all $N = 1.5 \times 10^5$ atoms, having mass $m = 1.445 \times 10^{-22}$ g and 
scattering length $a_s = 0.577 \times 10^{-6}$ cm, are Bose-condensed in a cylindrical 
harmonic trap with a radial frequency of $\om_x = 2\pi\times 210$ Hz and an axial 
frequency of $\om_z = 2\pi\times 23$ Hz. Hence the aspect ratio is 
$\nu = \omega_z/\omega_x = 0.11$. In experiments, the trapped atomic cloud had a radius 
of $R = 4 \times 10^{-4}$ cm and a length of $L = 6 \times 10^{-3}$ cm. In the 
Thomas-Fermi approximation, the cloud radius and length are
$$
R_{TF} ~ = ~ 1.036\; l_x(g\nu)^{1/5} \; ,
\qquad
L_{TF} ~ = ~ \frac{2.072}{\nu} \; l_x(g\nu)^{1/5} \;   ,
$$
while the oscillator lengths are $l_x = 0.74 \times 10^{-4}$ cm and 
$l_z = 2.25 \times 10 ^{-4}$ cm. The central density is 
$\rho=2.821 \times 10^{14}$ cm$^{-3}$. The coherence length, or healing length, is 
$$
 l_{coh} ~ = ~ \frac{\hbar}{mc} ~ = ~ \frac{1}{\sqrt{4\pi\rho a_s} } ~ = ~
1.37 \; \frac{l_x}{(g\nu)^{1/5}} \;  ,   
$$
which makes $l_{coh} = 2.18 \times 10^{-5}$ cm. The effective coupling parameter (\ref{35}) 
is $g = 1.96 \times 10^{4}$.

\section{Nonequilibrium states}

In the process of generation of nonlinear modes by a strong nonresonant modulation field, 
a whole sequence of nonequilibrium states has been observed in experiments 
\cite{Henn_139,Shiozaki_140,Seman_141,Bagnato_142} and modeled in computer simulations 
\cite{Yukalov_143,Yukalov_144,Yukalov_145,Novikov_146,Yukalov_147,Yukalov_148,Yukalov_149}.
Initially, elementary collective modes populate the system, followed by the production 
of vortices and solitons, which quickly begin to react, generating a cloud with the 
presence of large density fluctuations. The appearance of strong density fluctuations in 
the atomic cloud has been observed by subtracting from the current density distribution 
the averaged density, thus leaving only the strong fluctuations, as is shown in Fig. 4.
When we vary the amplitude for short excitation times, where vortices can be seen, we 
notice that the formation of visible vortices grows with the amplitude and with the 
excitation time. For sufficiently large amplitudes and excitation times, the evolution 
of the momentum distribution, projected onto the observation plane, evolves towards the 
occupation of the largest moments. The presence of just a few vortices does not seem to 
produce large variations in the density distribution, when compared with the excitation-free 
case. Passing through the turbulence threshold, the density distribution becomes displaced 
to the side of atoms with greater moments, and there appear the regions of momenta with a 
power law dependence in the distribution of moments, shown in Fig. 5. This is a signature 
of the presence of a direct cascade of particles, which is a characteristic feature of 
turbulence.

\begin{figure}[ht]
\centerline{
\includegraphics[width=8cm]{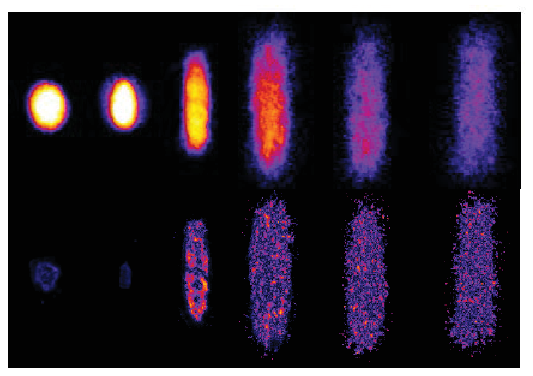} }
\caption{\small
Aspects of the excited cloud during free expansion and the distribution of the 
fluctuations obtained by subtracting the averaged profile.
}
\label{fig:Fig.4}
\end{figure}

\begin{figure}[ht]
\centerline{
\includegraphics[width=8cm]{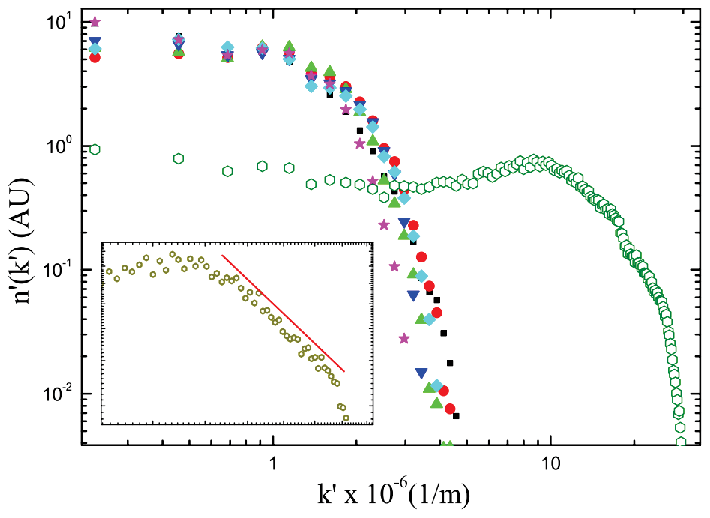} }
\caption{\small
Evolution of the projected momentum distribution measured by time of flight.
}
\label{fig:Fig.5}
\end{figure}

In more detail, the studied sequence of nonequilibrium states, observed in experiments
\cite{Henn_139,Shiozaki_140,Seman_141,Bagnato_142} and modeled in computer simulations 
\cite{Yukalov_143,Yukalov_144,Yukalov_145,Novikov_146,Yukalov_147,Yukalov_148,Yukalov_149}
is discussed below. 

\subsection{Weak nonequilibrium}

At the beginning of the process of imposing an alternating potential, when the injected 
energy $E_{inj}$ is yet insufficient for the creation of coherent modes, being less than 
twice the energy $E_{emb}$ required for creating at least a couple of embryos, or germs, 
of vortex rings,  
\be
\label{192}
0 ~ < ~ E_{inj} ~ < ~ 2E_{emb} \;   ,
\ee
the trapped atomic cloud is only slightly perturbed with small density fluctuations.
Vortex embryos appear by pairs, since each of them possesses vorticity $\pm 1$, while 
the vorticity of the whole system is zero, which requires that each vortex has to be 
compensated by an antivortex. The energy of creating a vortex embryo, $E_{emb}$, is 
smaller then that sufficient for generating a vortex ring, $E_{ring}$. For estimates, 
it is possible to accept that $E_{emb} \sim E_{ring}/2$. 

Let us bring attention of the reader that the injected energy (\ref{190}) is defined 
as the energy per particle. Therefore to compare it with the energies of coherent modes,
the corresponding energies below are also defined as characteristic reduced energies, 
that is the energies per particle. 

\subsection{Vortex embryos}

When the injected energy reaches the level sufficient for generating a pair of vortex
embryos, these objects arise being just pieces of broken vortex rings. This happens
when the injected energy is in the interval
\be
\label{193}
 2E_{emb}  ~ < ~ E_{inj} ~ < ~ 2E_{ring} \;   .
\ee
The embryos can be of different sizes. They have vorticity $\pm 1$. If the external 
pumping is switched off after the germs are created, they survive for about $0.2$ s, 
when they do not move, exhibiting only small oscillations. The local equilibration 
time is
$$
t_{loc} ~ = ~ \frac{m}{\hbar\rho a_s} ~ = ~ \frac{87.8}{(g\nu)^{2/5} \om_x} \; ,
$$
which is $t_{loc} \approx 10^{-3}$ s. Since the embryos lifetime of $0.2$ s is much 
larger than the local equilibration time, the embryos are metastable objects. Typical 
vortex embryos are shown in Fig. 6 obtained from numerical simulation. 

\begin{figure}[ht]
\centerline{
\includegraphics[width=9cm]{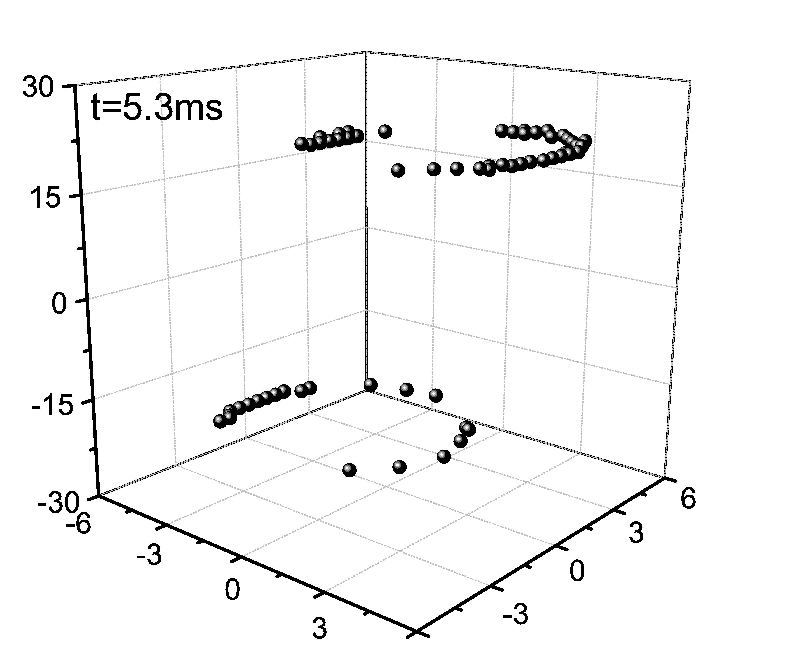} }
\caption{\small
Typical example of vortex germs. The time of the trap modulation is 
$t=5.3$ ms.
}
\label{fig:Fig.6}
\end{figure}

\subsection{Vortex rings}

The following pumping leads to the appearance of well formed rings of different radii 
$R_{ring}$, as in Fig. 7 representing numerical results. A vortex ring is a circular 
line of zero density, with a winding number $\pm 1$ around each of the line elements 
\cite{Iordanskii_150,Amit_151,Roberts_152,Jones_153,Barenghi_154,Belyaev_155}. The 
rings with opposite vorticity appear by pairs. These ring modes exist in the interval 
of energies
\be
\label{194}
  2E_{ring}  ~ < ~ E_{inj} ~ < ~ 2E_{vort} \;  ,
\ee
where $E_{vort}$ is the energy of a vortex line. The reduced energy of a ring is
\be
\label{195}
 E_{ring} ~ = ~ \frac{2\pi^2}{N} \; \rho \; R_{ring} \; l_x^2 \; \ln
\left( 2.25 \; \frac{R_{ring}}{l_{coh}} \right) \; \om_x \;  . 
\ee
Here $R_{ring}$ is the ring radius, $l_{coh}$ is the coherence length, and $\rho$ is 
the density of the system at the location of the ring, if the ring would be absent. 
In the trap, the ring oscillates \cite{Jackson_156,Reichl_157} with the period
\be
\label{196}
T_{ring} ~ = ~ \frac{4\pi R}{l_{coh}\om_z\sqrt{\ln(R/l_{coh})} } ~ = ~
\frac{4\pi(g\nu)^{2/5}}{\om_z\sqrt{\ln(0.5g\nu)} } \;   ,
\ee
which, for the accepted parameters, makes $T_{ring} \approx 0.7$ s. If the pumping 
is switched off, the ring lifetime is about $0.1$ s, which is much longer than the 
local equilibration time $t_{loc} \approx 10^{-3}$ s. This shows that the rings are 
metastable objects. In the trap, the rings look as practically immovable. A pair of
rings, found in numerical simulation, is shown in Fig. 7.

\begin{figure}[ht]
\centerline{
\includegraphics[width=9cm]{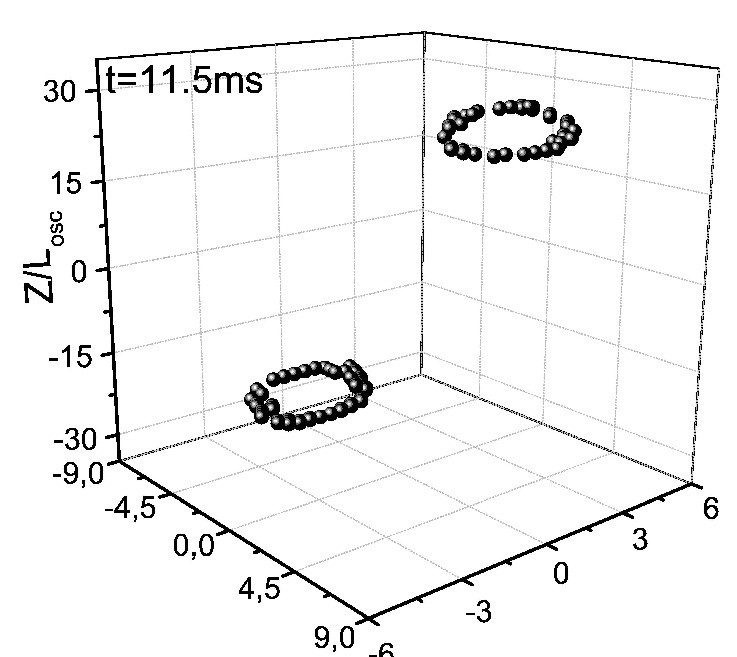} }
\caption{\small
Typical example of vortex rings. The time of the trap modulation is 
$t=11.5$ ms.
}
\label{fig:Fig.7}
\end{figure}

\subsection{Vortex lines}

The continuing pumping leads to the generation of pairs of vortex lines in the energy 
interval
\be
\label{197}
  2E_{vort}  ~ < ~ E_{inj} ~ < ~ E_{turb} \;   ,
\ee
in which $E_{turb}$ is the pumped energy after which vortex turbulence develops. Vortices
can have different lengths $L_{vort}$. The reduced energy of a vortex of length $L_{vort}$, 
in the Thomas-Fermi approximation \cite{Pethick_158}, reads as
\be
\label{198}
  E_{vort} ~ = ~ \frac{2\pi}{3N} \; \rho \; L_{vort} \; l_x^2 \; \ln
\left( 0.95 \; \frac{R}{l_{coh}} \right) \; \om_x \;  .
\ee
The vortex length $L_{vort}$ is approximately about the cloud length $L$. Then the reduced 
vortex energy is $E_{vort} = 0.296 \omega_x$. 

The vortices move randomly due to the Magnus force caused by the difference in 
pressure between the opposite sides of a vortex line. The vortex lifetime, after 
switching off the pumping, is about $0.2$ s. Numerically found vortex lines are 
shown in Fig. 8.

\begin{figure}[ht]
\centerline{
\includegraphics[width=9cm]{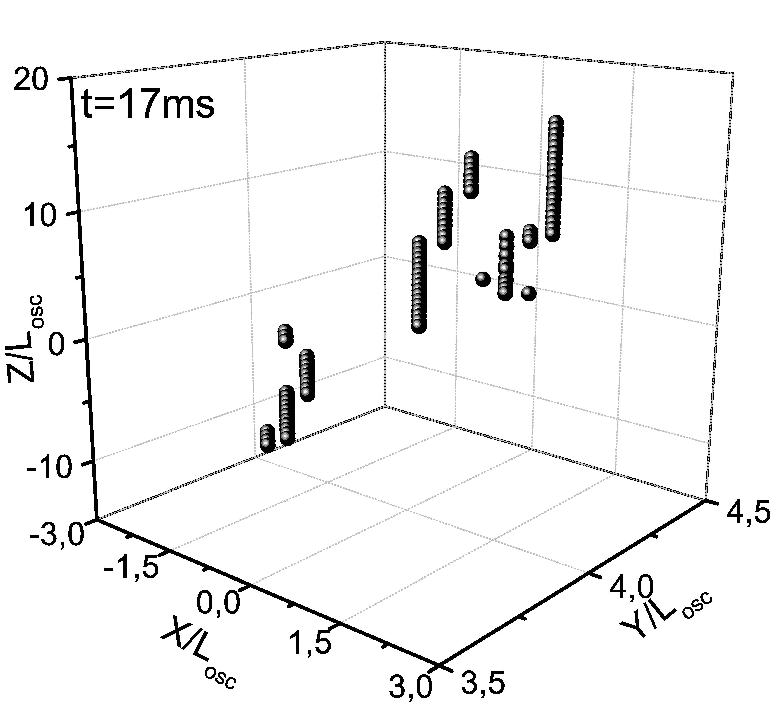} }
\caption{\small
Typical example of vortex lines directed along the $z$ 
axis. The time of the trap modulation is $t=17$ ms.
}
\label{fig:Fig.8}
\end{figure}

\subsection{Vortex turbulence}

Injecting into the trap more energy creates more and more vortices and anti-vortices, 
as well as distorts their straight lines and forces them to incline in different 
directions. The used modulating potential does not impose a rotation axis, because of 
which the increasing number of vortices does not lead to the creation of a vortex 
lattice, but results in the formation of a random vortex tangle corresponding to the 
definition of Vinen quantum turbulence 
\cite{Vinen_159,Vinen_160,Vinen_161,Tsubota_162,Nemirovskii_163,Tsatsos_2016}. Being 
released from the trap, the atomic cloud expands isotropically, which is typical of 
Vinen turbulence \cite{Henn_139,Shiozaki_140,Seman_141,Caracanhas_164}. 
 
Vortex turbulence exists in the energy interval
\be
\label{199}
 E_{turb}  ~ < ~ E_{inj} ~ < ~ E_{drop} \;  ,
\ee
where $E_{drop}$ is the upper boundary of the vortex turbulence. 

One distinguishes {\it random vortex turbulence}, or Vinen turbulence, as opposed 
to the {\it correlated vortex turbulence}, or Kolmogorov turbulence \cite{Nemirovskii_163}.
In a Vinen vortex tangle, the intervortex distance is much larger than the coherence 
length. Therefore, the summary energy of vortices in the random Vinen tangle can be 
approximated by the product of the vortex number and a single vortex energy. Thus the 
vortex turbulence energy threshold 
\be
\label{200}
E_{turb}  ~ = ~ E_{vort} \; N_{vort} 
\ee
is reached after the injected energy creates the critical number of vortices $N_{vort}$ 
that is so large that the mean distance between the outer parts of the vortices becomes 
equal to their diameter $2 \xi$, so that no vortex can be inserted inbetween without 
touching the neighbors. This implies that in the cross-section of the trap the 
number $N_{vort}$ is such that the equality
$$
N_{vort} \pi (4\xi)^2 ~ = ~ \pi R^2
$$
holds true. Hence the critical number of vortices is
\be
\label{201}
N_{vort} ~ = ~ \left( \frac{R}{4\xi}\right)^2 \;   .
\ee
For the experiments with $^{87}$ Rb, the atomic cloud radius is $R\approx 4\times 10^{-4}$ 
cm and the coherence length is $\xi\approx 2\times 10^{-5}$ cm. Thus, the critical vortex 
number is $N_{vort} \approx 25$, which is in perfect agreement with both experiments and 
numerical simulations. 

The overwhelming majority of vortices usually have the lowest circulation number $\pm 1$.
In the case where there would arise vortex lines with different circulation, it is 
possible to represent the system as a mixture of fluids containing vortex lines with 
different vorticity \cite{Yukalov_2010}.    

The column integrated radial momentum distribution obeys a power law 
$n_r(k) \propto k^{-\gamma}$ in the range $\pi/\xi < k < 2\pi/\xi$, which is a key 
quantitative expectation for an isotropic turbulent cascade \cite{Zakharov_165}. This
power low has been observed \cite{Thompson_166,Navon_167} for turbulent atomic clouds 
of trapped atoms, with the slope characterized by $\gm\approx 2$.

It is interesting to note that there exists a similar effect in laser media, which is 
called optical turbulence or turbulent photon filamentation. In high Fresnel lasers 
and nonlinear media there appear bright filaments randomly distributed in the sample 
cross-section and not correlated with each other. The theory of the photon turbulence 
is developed in Refs. \cite{Yukalov_168,Yukalov_169,Yukalov_170}. This effect has been 
observed in many experiments, as can be inferred from review articles 
\cite{Akhmanov_171,Lugiato_172,Arecchi_173,Huyet_174,Huyet_175,Arecchi_176,Calderon_177,
Yukalov_178,Yukalov_179}.

\subsection{Droplet state}

Starting from an equilibrium state of a practically completely Bose-condensed system, 
and injecting energy into the trap by applying an alternating field, we produce more
and more excited condensate. In weak nonequilibrium, the whole atomic cloud is yet 
Bose-condensed. The appearing vortex excitations, such as vortex embryos, vortex rings, 
and vortex lines can be treated as germs, or nuclei, of uncondensed phase. The 
following pumping of energy into the trap little by little destroys the condensate.
When the injected energy reaches the threshold value $E_{drop}$, vortex turbulence 
transforms into a new state composed of coherent Bose-condensed droplets immersed into 
an incoherent surrounding.       

The state of coherent droplets is analogous to heterophase states where different 
thermodynamic phases are intermixed on mesoscopic scale 
\cite{Yukalov_180,Yukalov_181,Yukalov_182}. This state exists in the energy interval
\be
\label{202}
 E_{drop}  ~ < ~ E_{inj} ~ < ~ E_c \;   ,
\ee
where $E_c$ is the critical amount of energy after which Bose-Einstein condensate is
completely destroyed and the lower energy threshold can be estimated as
\be
\label{203}
E_{drop} ~ = ~ E_{vort} \; N^*_{vort} \;   .
\ee
Here $N_{vort}^*$ is the maximal number of vortices that do not yet essentially 
interact with each other, so that their centers are separated by the distance 
$d_{int}$, which defines the number of vortices
\be
\label{204}
 N_{vort}^* ~ = ~ \left( \frac{R}{d_{int} } \right)^2 \;  .
\ee
With $R\approx 4 \times 10^{-4}$ cm and $d_{int}\approx 0.7 \times 10^{-4}$ cm, we 
have $N_{vort}^* \approx 30$, which is in good agreement with experiments. Since 
$E_{vort} = 0.296 \omega_x$, the energy, after which the droplet state develops is 
$E_{drop} \approx 8.88 \omega_x$.    

The droplet state develops because the number of vortices becomes so large and their 
interactions so strong that vortices destroy each other. Numerical simulations show
that the grains have the sizes close to the coherence length $\xi \approx 10^{-5}$ cm. 
The phase inside a grain is constant, hence the grains represent coherent droplets. 
After switching off the pumping by an alternating potential, each grain survives 
during the lifetime $10^{-2}$ s, which is much longer than the local equilibration time 
of $10^{-4}$ s. This tells us that, the droplets are metastable objects. The density 
inside a droplet is up to $100$ times larger than that of their incoherent surrounding. 
The number of atoms in a droplet is around $40$. The coherent droplets, whose number 
is about $400$, are randomly distributed inside the trap.  

Figure 9 shows experimentally measured density distributions for different states: 
vortex state with several vortices (a); vortex turbulence (b); and droplet state (c).

\begin{figure}[ht]
\centerline{
\includegraphics[width=12cm]{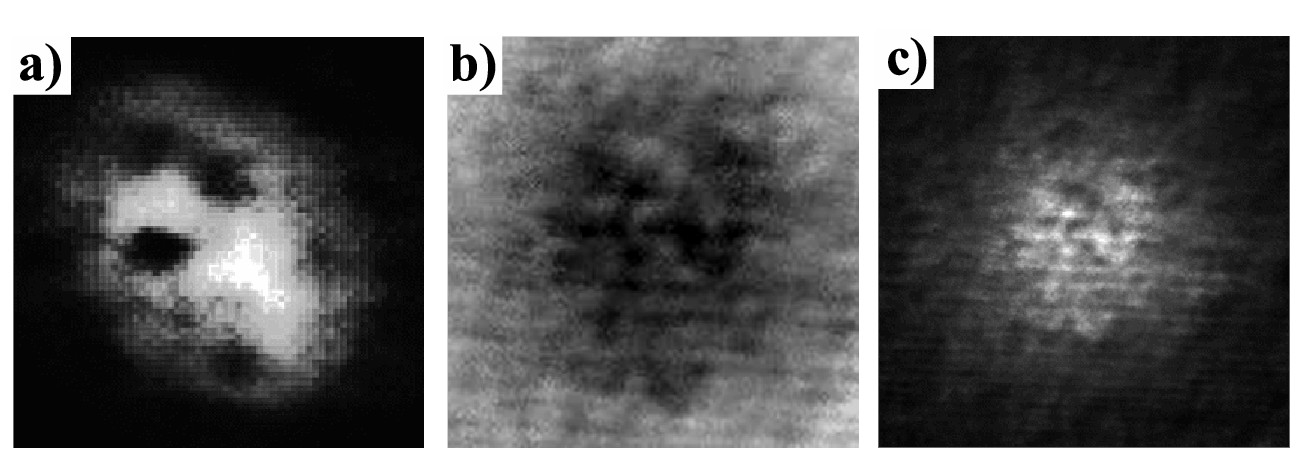} }
\caption{\small
 Density distributions for different states, observed in
experiments:
(a) vortex state with several vortices; (b) vortex turbulence; (c) droplet
state.
}
\label{fig:Fig.9}
\end{figure}

The droplet state can be called heterophase droplet mixture, droplet turbulence, or
grain turbulence. It also reminds fog. In numerical simulations, considering the 
opposite process of equilibrium Bose-condensate formation from an uncondensed gas, 
this regime is termed strong turbulence \cite{Zakharov_165,Zakharov_183}. Since the 
location of droplets in space is random, after averaging over spatial configurations, 
the system looks as a mesoscopic mixture of two phases with different density
\cite{Yukalov_184,Yukalov_185}. It is possible to show that the averaging over phase 
configurations is similar to the averaging over time \cite{Yukalov_186}.

\subsection{Wave turbulence}

After the energy injected into the trap reaches the value $E_c = k_B T_c$, with $T_c$ 
being the Bose-Einstein condensation temperature, the condensed state cannot exist. 
Then there develops a state modeling the turbulent normal state. This state corresponds
to what is termed wave turbulence or weak turbulence. These waves are not so big, with 
the density only about $3$ times larger than their surrounding. The typical size of a 
wave is of order $10^{-4}$ cm. The phase inside it is random, which is in line with the
absence of Bose-Einstein condensate. However, strictly speaking, the transition between 
the droplet state and wave turbulence is a gradual crossover. The crossover line, 
conditionally separating the regimes of droplet state and wave turbulence, can be 
located at the point, where the system coherence, associated with droplets, is destroyed 
in half of the system. For $^{87}$Rb atoms, the critical temperature is 
$T_c = 2.76 \times 10^{-7}$ K, hence $E_c = 27.4 \omega_x$.

\section{Classification of states}

The described nonequilibrium states can be distinguished by the intervals of injected 
energy, keeping in mind the relations $E_{emb} \approx 0.5 E_{ring}$ and 
$E_{ring} \approx \nu E_{vort}$, with $\nu = 0.11$. Then the characteristic energies are:
$$
E_{emb} ~ = ~ 0.016 \om_x \; , \qquad E_{ring} ~ = ~ 0.033 \om_x \; ,
$$
$$
E_{vort} ~ = ~ 0.296 \om_x \; , \qquad E_{turb} ~ = ~ 7.4 \om_x \; ,
$$
\be
\label{205}
 E_{drop} ~ = ~ 8.88 \om_x \; , \qquad E_c ~ = ~ 27.4 \om_x \;   .
\ee
Introducing the dimensionless injected energy $E \equiv E_{inj}/\omega_x$, it is possible 
to list the related energy intervals as follows:
\begin{eqnarray}
\nonumber
\begin{array}{rlllll}
0     & < & E & < &  0.032 ~~~ & ~~~ ( weak \; nonequilibrium) \; , \\
0.032 & < & E & < &  0.066 ~~~ & ~~~ ( vortex \; embryos) \; , \\
0.066 & < & E & < &  0.592 ~~~ & ~~~ ( vortex \; rings) \; , \\
0.592 & < & E & < &   7.4  ~~~ & ~~~ ( vortex \; lines) \; , \\
7.4   & < & E & < &  8.88  ~~~ & ~~~ ( vortex \; turbulence) \; , \\
8.88  & < & E & < &  27.4  ~~~ & ~~~ ( coherent \; droplets) \; , \\
      &   & E & > &  27.4  ~~~ & ~~~ ( wave \; turbulence) \; .
\end{array}
\end{eqnarray}

When a system, prepared in a strongly nonequilibrium symmetric state, equilibrates 
to a state with broken symmetry, there appear topological defects, such as vortices, 
protodomains, cells, grains, strings, and monopoles. This is called the Kibble-Zurek 
mechanism \cite{Kibble_187,Kibble_188,Zurek_189,Zurek_190}. In our experiments, the 
opposite procedure of generating nonequilibrium states with topological modes is 
realized, which can be called {\it inverse Kibble-Zurek scenario} \cite{Yukalov_145}. 

The nonequilibrium states can also be characterized \cite{Yukalov_148,Yukalov_149} 
by effective temperature, Fresnel, and Mach numbers. The {\it effective temperature}
is given by the relation
\be
\label{206}
T_{eff} ~ \equiv ~ \frac{2}{3} \; [\; E_{kin}(t) - E_{kin}(0) \; ] \;  ,
\ee 
where $E_{kin}(t)$ is the current kinetic energy per particle and $E_{kin}(0)$ is the
initial kinetic energy of the system at equilibrium. The Fresnel number can be defined
as
\be
\label{207}
 F ~ \equiv ~ \frac{\pi R^2}{\lbd L}  \qquad 
\left( \lbd \equiv \sqrt{ \frac{2\pi}{m T_{eff}} } \right) \; ,
\ee
with $\lambda$ being thermal wavelength. And the Mach number is written as
\be
\label{208}
 M ~ \equiv ~ \frac{v}{c} ~ = ~ \sqrt{ \frac{3T_{eff}}{mc^2} } \;  ,
\ee
where $v$ is the particle velocity corresponding to the kinetic energy and $c$ is 
the sound velocity in the system. The classification of the nonequilibrium states of 
trapped $^{87}Rb$ is summarized in Table 1. The values of effective temperature, Fresnel
number and Mach number correspond to the lower boundary of the related states, where
these states appear.

\begin{table}[hp]
\caption{\small Nonequilibrium states of a trapped Bose-Einstein condensate, 
characterized by effective temperature (in units of $\om_x$), Fresnel number, 
and mach number. The figures correspond to the lower boundary of the related 
states.}

\vskip 3mm
\centering
\renewcommand{\arraystretch}{1.2}
\begin{tabular}{|c|c|c|c|} \hline
Nonequilibrium states & Effective temperature & Fresnel number & Mach number  \\ \hline
wave turbulence       & 23.5                  & 1.01           & 1.07   \\ 
coherent droplets     & 8.56                  & 0.61           & 0.66    \\ 
vortex turbulence     & 5.54                  & 0.49           & 0.53  \\ 
vortex lines          & 2.26                  & 0.31           & 0.25  \\
vortex rings          & 1.21                  & 0.23           & 0.22   \\ 
vortex embryos        & 0.29                  & 0.11           & 0.08  \\ \hline
weak nonequilibrium   &                       &                &  \\ \hline
\end{tabular}
\end{table}

\section{Atom laser}

If trapped atoms could be ejected from the trap, this would remind the action of a laser
emitting photons. Bose-Einstein condensate is a coherent system, hence emitting 
Bose-condensed atoms would be similar to emitting coherent light. Thus, one compares
with the optical laser the simple process of Bose atoms falling out of the trap under
the influence of gravity \cite{Mewes_191,Bloch_192,Bolpassi_193,Bennets_194,Chen_195}.
The direction of an atomic beam can be regulated by means of Raman scattering 
\cite{Hagley_196}. A mechanism allowing for guiding atomic beams in any required 
direction has been proposed and analyzed in Refs. 
\cite{Yukalov_197,Yukalov_198,Yukalov_199,Yukalov_200}.

Let us consider a system of atoms, with spin operators $S_i$, with interaction potential 
$\Phi_{ij}$, in a magnetic field ${\bf B}_i$, and subject to the Earth gravity force 
${\bf g}$. The Hamiltonian is
\be
\label{209}
 H ~ = ~ \sum_{i=1}^N \left( -\; 
\frac{\nabla_i^2}{2m} \; - \; \mu_S\; \bS_i \cdot\bB_i - m\; {\bf g} \cdot \br_i 
\right) + \frac{1}{2} \sum_{i\neq j} \Phi_{ij} \; .
\ee
Here $\mu_S$ is the atomic magnetic moment that can be positive or negative. Since 
magnetic moments of atoms are often mainly due to electrons, for concreteness, we assume 
that $\mu_S < 0$, although the results do not principally depend on the sign of the 
magnetic moment. 

Denoting the quantum-mechanical average by angular brackets, it is straightforward to derive
the evolution equations for the mean position vector,
\be
\label{210}
 m\; \frac{d^2}{dt} \; \lgl \; \br_i \; \rgl ~ = ~ 
\mu_S \lgl \; \vec{\nabla}_i (\bS_i \cdot \bB_i) \; \rgl - 
\sum_{j(\neq i)}^N \lgl \; \vec{\nabla}_i \; \Phi_{ij}\; \rgl + m\; {\bf g} \; ,
\ee
and the average spin,
\be   
\label{211}
 \frac{d}{dt}\; \lgl \; \bS_i \; \rgl ~ = ~ 
\mu_S \; \lgl \; \bS_i \times \bB_i \; \rgl \;  .
\ee

Being interested in the motion of an atomic cloud as a whole, we introduce the 
center-of-mass vector
\be
\label{212}
 {\bf R} ~ \equiv ~ \frac{1}{N} \sum_{i=1}^N \lgl \; \br_i \; \rgl \;  ,
\ee
the average spin
\be
\label{213}
\bS ~ \equiv ~ \; \frac{1}{N} \sum_{i=1}^N \lgl \; \bS_i \; \rgl \;  ,
\ee
and the collision force
\be
\label{214}
 {\bf I} ~ \equiv ~ - \; 
\frac{1}{N} \sum_{i\neq j}^N \lgl \; \vec{\nabla}_i \; \Phi_{ij} \; \rgl 
\ee
due to atomic interactions.

Assume that the magnetic field is slowly varying in space, so that
\be
\label{215}
\bB_i ~ \equiv \bB(\br_i) ~ \cong~ \bB({\bf R}) ~ \equiv ~ \bB    .
\ee
Then the approximate equalities are valid:
$$
\frac{1}{N} \sum_{i=1}^N  \lgl \; \vec{\nabla_i} (\bS_i \cdot \bB_i) \; \rgl ~
\cong ~ \vec{\nabla} (\bS \cdot \bB) \; ,
$$
$$
\frac{1}{N} \sum_{i=1}^N  \lgl \; \bS_i \times \bB_i \; \rgl ~
\cong ~ \bS \times \bB \; .
$$
Therefore, we have the equations of motion for the center of mass
\be
\label{216}
m\; \frac{d^2{\bf R}}{dt^2} ~ = ~ 
\mu_S \vec{\nabla}(\bS \cdot\bB) + {\bf I} + m{\bf g} 
\ee
and for the average spin
\be
\label{217}
 \frac{d\bS}{dt} ~ = ~ \mu_S \; \bS \times \bB \;   .
\ee
 
The magnetic field of the trap can be taken in the typical of magnetic traps form
\cite{Petrich_201,Anderson_202,Jin_203}, as the sum 
\be
\label{218}
\bB ~ = ~ \bB_1 + \bB_2
\ee
of a quadrupole field
\be
\label{219}
\bB_1 ~ = ~ B_1' \; ( R_x \bfe_x + R_y \bfe_y - \lbd R_z \bfe_z ) \qquad ( B_1' > 0 )
\ee
and a transverse field
\be
\label{220}
 \bB_2 ~ = ~ B_2 (h_x \bfe_x + h_y \bfe_e ) \qquad ( B_2 > 0 ) \;  ,
\ee
where
$$
h_x^2 + h_y^2 ~ = ~ 1 \qquad (B_2 = |\; \bB_2\; | ) \; .
$$
There exist two types of quadrupole traps, dynamic quadrupole traps, with the transverse 
field depending on time,
\be
\label{221}
 h_x ~ = ~ \cos(\om t) \; , \qquad h_y ~ = ~ \sin(\om t) \;  ,
\ee
and static quadrupole traps, with a time-independent transverse field,
\be
\label{222}
h_x ~ = ~ const \; , \qquad  h_y ~ = ~ const \;  .
\ee
Also, there can exist an additional so-called cooling field serving for removing fast 
atoms from the center of the trap. However this cooling field does not play role in
describing the mechanism of creating directed atomic beams \cite{Yukalov_198}.   
 
The method we are suggesting does not depend on whether a dynamic or static trap is 
accepted \cite{Yukalov_204,Yukalov_205}. For concreteness, below we illustrate the 
results for the dynamic quadrupole trap.  

The spatial motion of the center of mass is characterized by the frequency $\omega_1$
defined by the equation
\be
\label{223}
 \om_1^2 ~ = ~ - \; \frac{\mu_S(B_1')^2}{mB_2} ~ > ~ 0 \;  ,
\ee
while the spin motion is characterized by the Zeeman frequency
\be
\label{224}
\om_2 ~ = ~ - \mu_S B_2 ~ > ~ 0 \;   .
\ee
We stress that these frequencies are positive, keeping in mind that $\mu_S < 0$. 

The spin motion is faster than the spatial atomic motion, which implies that 
$\omega_2 \gg \omega_1$. The rotation of the transverse field is faster than the atomic 
motion, $\omega \gg \omega_1$, but slower than the spin motion in order not to induce 
transitions between the Zeeman energy levels, hence $\omega \ll \omega_2$. Thus there 
are the relations
\be
\label{225}
\om_1 ~ \ll ~ \om ~ \ll ~ \om_2 \;   .
\ee
 
The size of the atomic cloud is defined by the effective radius
\be
\label{226}
 R_0 ~ \equiv ~ \frac{B_2}{B_1'} \;  .
\ee
This radius can be used for defining dimensionless spatial variables describing the 
location of the center of mass,
\be
\label{227}
 \br ~ \equiv ~ \frac{{\bf R}}{R_0} ~ = ~ \{x,y,z\} \;  .
\ee

Let us introduce the collision rate $\gamma$ by the equality
\be
\label{228}
\gm\vec{\xi} ~ \equiv ~ \frac{{\bf I}}{mR_0} \;   .
\ee
The cloud of trapped atoms is rarified, so that the collision rate is small, in the 
sense that
\be
\label{229}
\gm ~ \ll ~ \om_1 \;   ,
\ee
and $\xi$ can be approximated by a random variable centered at zero because of the random
directions of gradients in the collision force (\ref{214}).  

The shape of the trap can be described by a shape factor, which can be defined by the 
function
\be
\label{230}
f(\br) ~ = ~ \exp\left\{ -\; \frac{x^2+y^2}{(R/R_0)^2} \; - \; 
\frac{z^2}{(L/R_0)^2} \right\} \;   ,
\ee
where the trap radius $R \geq R_0$ and the trap length $L \geq R_0$, or by a step-like 
function
\be
\label{231}
 f(\br) ~ = ~ 1 - \Theta\left( x^2 + y^2 \; - \; \frac{R^2}{R_0^2} \right) \;
\Theta\left( |\; z\;| - \; \frac{L}{2R_0} \right) \;  .
\ee
As we have checked, the concrete choice of the shape factor does not change the overall 
picture and does not influence the described mechanism \cite{Yukalov_204,Yukalov_205}.  

Thus, the center-of-mass spatial motion is given by the equation
\be
\label{232}
\frac{d^2\br}{dt^2} ~ = ~ \left(\; {\bf F} + \gm \vec{\xi} \; \right) \; f +
 \frac{{\bf g}}{R_0} \;   ,
\ee
with $f = f({\bf r})$ and the force
\be
\label{233}
{\bf F} ~ = ~ - \om_1 \; ( S_x\bfe_x + S_y \bfe_y - \lbd S_z \bfe_z ) \; .
\ee
 
The spin motion, characterized by equation (\ref{217}), can be represented as
\be
\label{234}
\frac{d\bS}{dt} ~ = ~ \om_2 \hat A \bS \;   ,
\ee
with an antisymmetric matrix $\hat{A}=[A_{\alpha \beta}]$ having the elements
$$
A_{\al\al} ~ = ~ 0 \; , \qquad A_{\al\bt} ~ = ~ -A_{\bt\al} \; , 
\qquad A_{21} = \lbd\; z \; ,
$$
\be
\label{235}
A_{23} ~ = ~ x + h_x \; , \qquad A_{31} ~ = ~ y +h_y \; .
\ee

According to condition (\ref{225}), the transverse field (\ref{221}), rotating with 
the frequency $\omega$, is fast, as compared to the atomic motion characterized by 
the frequency $\omega_1$. Then, it is possible to resort to the averaging method
\cite{Bogolubov_206}, following which, we find the solution to the fast variable, 
that is the spin ${\bf S}$, keeping the slow variables, that is the atomic coordinates
${\bf r}$, fixed, and substitute the found fast solution to the equation for the slow
variable (\ref{232}), with averaging the right-hand side of this equation over the 
random variable $\vec{\xi}$ and over time (see details in Refs. 
\cite{Yukalov_207,Yukalov_208,Yukalov_209}). Thus we come to the equation
\be
\label{236}
\frac{d^2\br}{dt^2} ~ = ~ {\bf F}_{eff} \; f + \frac{{\bf g}}{R_0} \;   ,
\ee
where $f = f({\bf r})$ and the effective force is
\be
\label{237}
{\bf F}_{eff}  ~ = ~ -\;
\frac{\om_1^2[\;(1+x)S_0^x+yS_0^y-\lbd z S_0^z\;](x\bfe_x+y\bfe_y+2\lbd^2z\bfe_z)}
{2[\;(1+2x+x^2+y^2+\lbd^2 z^2)(1+x^2+y^2+\lbd^2 z^2)\;]^{1/2} } \; .
\ee
Here $S_0^\alpha$ is the initial spin polarization.     

The standard initial spin polarization is $S_0^x = S > 0$ and $S_0^y = S_0^z = 0$. This
guarantees atomic confinement for $r < 1$, with the attractive effective force
$$
{\bf F}_{eff}  ~ \cong ~ -\; 
\frac{1}{2} \; \om_1^2 \; S \; ( x\bfe_x+y\bfe_y+2\lbd^2 z \bfe_z ) \; ,
$$
leading to the effective potential well
$$
U_{eff} ~ \cong ~ \frac{1}{4} \; \om_1^2 \; S \;
(x^2 + y^2 + 2 \lbd^2 z^2)  \; .
$$

On the contrary, we suggest to arrange the initial spin polarization under conditions
\be
\label{238}
S_0^x ~ = ~ S_0^y ~ = ~ 0 \; , \qquad S_0^z ~ = ~ S \;   ,
\ee 
which results in the effective force
\be
\label{239}
{\bf F}_{eff}  ~ = ~ 
\frac{\om_1^2\lbd S z (x\bfe_x+y\bfe_y+2\lbd^2z\bfe_z)}
{2[\;(1+2x+x^2+y^2+\lbd^2 z^2)(1+x^2+y^2+\lbd^2 z^2)\;]^{1/2} } \;   .
\ee
If the spin polarization has been aligned with the axis $x$, then the desired initial
spin polarization (\ref{238}) along the axis $z$ can be prepared \cite{Slichter_210} 
by the action of an oscillating $\pi/2$ pulse with the frequency in resonance with 
$\omega_2$, turning the spin from the axis $x$ to the axis $z$.  

Let us denote a dimensionless gravity acceleration 
\be
\label{240}
\vec{\dlt} ~ \equiv ~ \frac{{\bf g}}{R_0 \lbd S \om_1^2} ~ = ~ 
\frac{m{\bf g}}{|\; \lbd S \mu_s B_1'\;|} \;   .
\ee
Then equations (\ref{236}) acquire the form
$$
\frac{d^2x}{dt^2} ~ = ~ \frac{1}{2} \; z \; x\; f + \dlt_x \; ,
$$
\be
\label{241}
\frac{d^2y}{dt^2} ~ = ~ \frac{1}{2} \; z \; y\; f + \dlt_y \; , \qquad
 \frac{d^2z}{dt^2} ~ = ~ \lbd^2\; z^2 \; f + \dlt_z \;  .
\ee

The detailed analysis of these equations at small $x$, $y$, and $z$, as well as numerical 
solution for arbitrary spatial coordinates, demonstrate the effect of {\it atomic 
semiconfinement} \cite{Yukalov_197,Yukalov_198,Yukalov_199,Yukalov_200,Yukalov_204,
Yukalov_205,Yukalov_207,Yukalov_208,Yukalov_209}, when atoms are confined from one side
of the trap, but are expelled from another side. In the present setup, atoms are ejected
in the direction of the trap axis, with $z > 0$. The typical trajectories of the emitted 
atomic beams are shown in Fig. 10 for the trap axis $z$ inclined by $45$ degrees to 
the horizon, so that $\delta_z = - \delta_x$; the initial spatial location of the atomic 
cloud center of mass is close to the trap center and the initial dimensionless velocities 
are close to zero, being varied in the interval $-0.1 \leq v_0^\alpha < 0.1$. The aspect 
ratio $\nu = <x>^2/<z>^2$ is rather small, being $10^{-4}$, which means that the atomic 
beam is stretched in the $z$ direction about $100$ times larger then in the $x$ direction, 
that is the beam is very well collimated. The collimation is not essentially influenced 
by thermal fluctuations, provided the temperature is less than the critical temperature 
$$
 T_c ~ \approx ~ \frac{\hbar \om_1^3}{k_B \gm^2} \;  ,
$$
which is close to the Bose condensation temperature. After leaving the trap, the 
trajectories become curved due to gravity. The velocity of emitted atoms depends on 
the setup parameters. For realistic parameters, the velocity can reach $100$ cm/s.

\begin{figure}[ht]
\centerline{
\hbox{ \includegraphics[width=7.5cm]{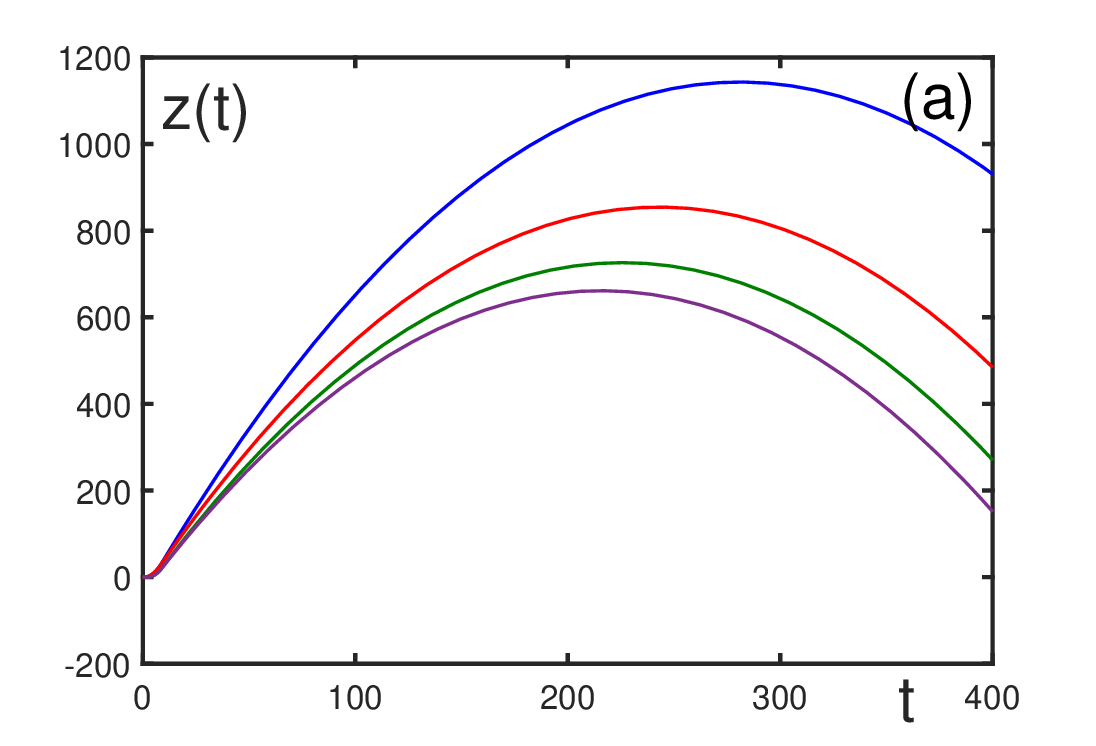} \hspace{1cm}
\includegraphics[width=7.5cm]{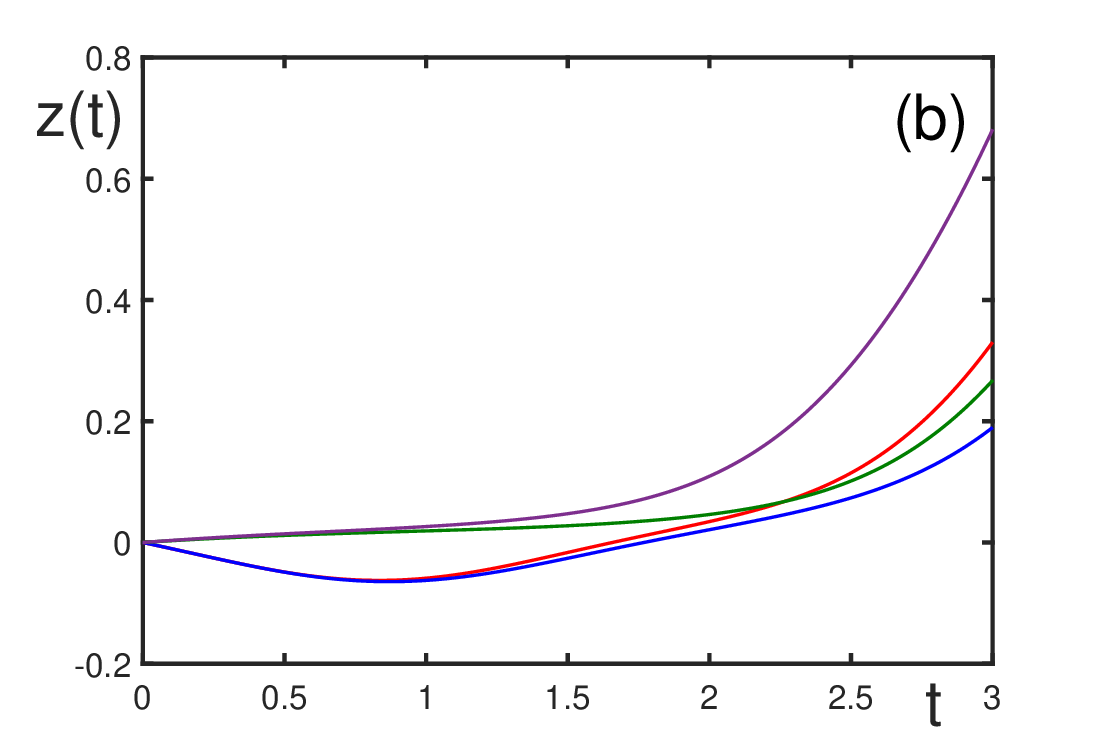}  } }
\caption{\small
Typical trajectories for the trap parameters 
$L=40$, $R=1$, $\lbd=10$, the gravity components $\dlt_x=0.03$, $\dlt_z=-\dlt_x$, 
$\dlt_y=0$, and different velocities $-0.1<\dot{r}<0.1$:
(a) for the time interval $0\leq t\leq 400$;
(b) beginning of the process;
}
\label{fig:Fig.10}
\end{figure}

\section{Conclusion}

The problem of generating non-ground-state Bose-Einstein condensates is discussed. 
Such non-ground-state Bose condensates are represented by nonlinear coherent modes
of trapped atoms. Coherent modes are the solutions to the stationary nonlinear 
Schr\"{o}dinger equation. The solution to the time-dependent nonlinear Schr\"{o}dinger 
equation can be expanded over the coherent modes, which leads to the equations for
the amplitudes of the fractional mode populations. 

The generation of the coherent modes can be realized by alternating external fields,
when either an external trapping potential or atomic interactions are modulated. There
exist two ways of generation, resonant generation and nonresonant generation. In the
resonant generation, the perturbing alternating field does not need to be strong, but 
the alternating frequency has to be in resonance with the transition frequency 
corresponding to the difference between the energy levels of the ground and excited 
modes. In the nonresonant generation, the perturbing field has to be sufficiently 
strong, however there is no need to tune the frequencies to resonant conditions.      

In the behaviour of coherent modes, there are many similarities with optical phenomena,
such as mode locking, interference patterns, Josephson current, Rabi oscillations,
Ramsey fringes, and the occurrence of higher-order resonances related to harmonic 
generation and parametric conversion. Dynamic transition between mode-locked and mode 
unlocked regimes is similar to a critical phase transition in a statistical system. 

Creating coherent modes in a system of coupled traps or in a deep optical lattice
allows for the realization of mesoscopic entanglement of atomic clouds and entanglement 
production. With coherent modes, one can observe atomic squeezing   

Employing nonresonant generation it is possible to form strongly nonequilibrium states 
of Bose condensates with different topological modes including vortex embryos, vortex 
rings, vortex lines, and condensate droplets. Such nonequilibrium states as vortex 
turbulence, droplet turbulence, and wave turbulence can be produced. The nonequilibrium 
states can be classified by means of effective temperature, Fresnel and Mach numbers.  
The overall procedure of perturbing an initially equilibrium condensate to nonequilibrium 
states with topological defects, such as vortex embryos, vortex rings, vortex lines, 
and condensate droplets is called inverse Kibble-Zurek scenario. 

The possibility of producing well collimated atomic beams from atom lasers, which can be
guided in any desired direction is described.  

\vskip 2cm

This research did not receive any specific grant from funding agencies in the public, 
commercial, or not-for-profit sectors.

\end{document}